\documentclass[10pt]{article}
\usepackage{amssymb}
\usepackage{amsfonts}
\usepackage{amsmath}

\numberwithin{equation}{section}

\newcommand{\bR}{{\mathbb R}}

\newcommand{\bC}{{\mathbb C}}

\newcommand{\bN}{{\mathbb N}}

\newcommand{\kC}{{\mathcal C}}

\newcommand{\kH}{{\mathcal H}}

\newcommand{\kN}{{\mathcal N}}

\newcommand{\gotH}{{\mathfrak H}}

\newcommand{\gotK}{{\mathfrak K}}

\newcommand{\gotL}{{\mathfrak L}}
\newcommand{\gotm}{{\mathfrak m}}

\newcommand{\gga}{{\gamma}}
\newcommand{\gG}{{\Gamma}}
\newcommand{\gk}{{\kappa}}

\newcommand{\gl}{{\lambda}}

\newcommand{\gP}{{\Pi}}

\newcommand{\gS}{{\Sigma}}
\newcommand{\gs}{{\sigma}}
\newcommand{\gt}{{\tau}}

\newcommand{\gT}{{\Theta}}

\newcommand{\gY}{{\Upsilon}}

\newcommand{\slim}{\,\mbox{\rm s-}\hspace{-2pt} \lim}
\newcommand{\wlim}{\,\mbox{\rm w-}\hspace{-2pt} \lim}

\newcommand{\real}{{\Re{\mathrm e\,}}}
\newcommand{\imag}{{\Im{\mathrm m\,}}}
\newcommand{\dom}{{\mathrm{dom\,}}}
\newcommand{\ran}{{\mathrm{ran\,}}}

\newcommand{\mul}{{\mathrm{mul}}}
\newcommand{\op}{{\mathrm{op}}}

\newcommand{\clo}{{\mathrm{clo}}}

\newcommand{\spa}{{\mathrm{span}}}
\newcommand{\clospa}{{\mathrm{clospan}}}

\newtheorem{thm}{Theorem}[section]
\newtheorem{prop}[thm]{Proposition}
\newtheorem{lem}[thm]{Lemma}
\newtheorem{cor}[thm]{Corollary}

\newtheorem{defn}[thm]{Definition}
\newtheorem{rem}[thm]{Remark}

\newcommand{\ba}{\begin{array}}
\newcommand{\ea}{\end{array}}
\newcommand{\bea}{\begin{eqnarray}}
\newcommand{\eea}{\end{eqnarray}}
\newcommand{\bead}{\begin{eqnarray*}}
\newcommand{\eead}{\end{eqnarray*}}
\newcommand{\be}{\begin{equation}}
\newcommand{\ee}{\end{equation}}
\newcommand{\bed}{\begin{displaymath}}
\newcommand{\eed}{\end{displaymath}}
\newcommand{\bl}{\begin{lem}}
\newcommand{\el}{\end{lem}}
\newcommand{\bp}{\begin{prop}}
\newcommand{\ep}{\end{prop}}
\newcommand{\bt}{\begin{thm}}
\newcommand{\et}{\end{thm}}
\newcommand{\Label}{\label}
\newcommand{\bc}{\begin{cor}}
\newcommand{\ec}{\end{cor}}
\newcommand{\la}{\Label}

\newcommand{\br}{\begin{rem}}
\newcommand{\er}{\end{rem}}
\newcommand{\bd}{\begin{defn}}
\newcommand{\ed}{\end{defn}}

\newenvironment{proof}%
{\begin{sloppypar}\noindent{\bf Proof.}}%
{\hspace*{\fill}$\square$\end{sloppypar}\bigskip}

   \def\sH{{\mathfrak H}}   
   \def\sK{{\mathfrak K}}

      \def\dC{{\mathbb C}}

   \def\dN{{\mathbb N}}   
      \def\dR{{\mathbb R}}

      \def\cC{{\mathcal C}}
      
\def\cG{{\mathcal G}}   \def\cH{{\mathcal H}}   
      
   \def\cN{{\mathcal N}}

\def\mul{{\text{\rm mul\,}}}

\title{On Eisenbud's and Wigner's $R$-matrix: \\
A general approach}

\author{
J.~Behrndt$^{\;a}$,
H.~Neidhardt$^{\;b}$,
E.~R.~Racec$^{\;c}$,\\ 
P.~N.~Racec$^{\;d},$ 
U.~Wulf$^{\;e}$
}

\date{January 18, 2007}

\begin{document}

\maketitle 
\vspace{-0.5cm}
{\small
\begin{quote}

\begin{enumerate}

\item[a)] Technische Universit\"{a}t Berlin,
Institut f\"ur Mathematik, 
Stra\ss e des 17.\ Juni 136, D-10623 Berlin, Germany\\[1mm]
E-mail: behrndt@math.tu-berlin.de

\item[b)] Weierstra{\ss}-Institut f\"ur Angewandte Analysis und Stochastik,
Mohrenstr. 39, D-10117 Berlin, Germany\\[1mm]
E-mail: neidhard@wias-berlin.de

\item[c)] Technische Universit\"at Cottbus, Fakult\"at 1, Postfach 101344, 
 D-03013 Cottbus, Germany, and \\[1mm]
Faculty of Physics, University of Bucharest, PO Box MG-11,
077125 Bucharest Magurele, Romania \\[1mm]
E-mail: roxana@physik.tu-cottbus.de

\item[d)] Weierstra{\ss}-Institut f\"ur Angewandte Analysis und Stochastik,
Mohrenstr. 39, D-10117 Berlin, Germany, and \\[1mm]
National Institute of Materials Physics, PO Box MG-7,
077125 Bucharest Magurele, Romania \\[1mm]
E-mail: racec@wias-berlin.de

\item[e)] Technische Universit\"at Cottbus, Fakult\"at 1, Postfach 101344, 
D-03013 Cottbus, Germany\\[1mm]
E-mail: wulf@physik.tu-cottbus.de

\end{enumerate}
\end{quote}
}

\begin{abstract}
\noindent
The main objective of this paper is to give a rigorous treatment of Wigner's and Eisenbud's
$R$-matrix method for scattering matrices of 
scattering systems consisting of two selfadjoint extensions
of the same symmetric operator with finite deficiency indices. In the
framework of boundary triplets and associated Weyl functions an abstract generalization of the 
$R$-matrix method is developed and the results are applied 
to Schr\"odinger operators on the real axis.
\end{abstract}


\vspace{-3mm}

\section{Introduction}


The $R$-matrix approach to scattering was originally developed by Kapur
and Peierls \cite{KP1} in connection with nuclear reactions. Their ideas
were improved by Wigner \cite{W1,W0} and Wigner and Eisenbud
\cite{W2}, where the notion of the $R$-matrix firstly occurred.
A comprehensive overview of the $R$-matrix theory in nuclear
physics can be found in \cite{B1,LTh1}. 
The key ideas of the $R$-matrix theory are rather
independent from the concrete physical situation. In fact,
later the $R$-matrix
method has also found several applications in atomic and molecular physics
(see e.g. \cite{BB1,BR1})
and recently it was applied to transport problems 
in semiconductor nano-structures
\cite{NWR1,NWR2,OKW1,RRW2,RRW1,RW1,WKRS1,WKS1,WRRA1}. 
In \cite{N1,N2} an attempt was made to make the $R$-matrix method
rigorous for elliptic differential operators,
see also \cite{S1,S3} for Schr\"odinger operators and \cite{S4,S2} for an extension 
to Dirac operators.

The essential idea
of the $R$-matrix theory is to divide the whole physical system into two
spatially divided subsystems which are called internal and external
systems, see \cite{W1,W0,W2}. The internal system is usually related 
to a bounded region, while the external system is given on its complement
and is, therefore, spatially infinite. The goal is to represent the scattering matrix of a
certain scattering system
in terms of eigenvalues and eigenfunctions of an operator corresponding to the internal system with
suitable chosen selfadjoint boundary conditions at the interface
between the internal and external system. This might seem a little strange at first sight since
scattering is rather related to the external system than to the internal one.

It is the main objective of the present paper to make a further step towards a rigorous foundation of the
$R$-matrix method in the framework of abstract scattering theory
\cite{BW1}, in particular, in the framework of scattering theory for open quantum systems 
developed in \cite{BMN2,BMN1}. This abstract approach has the
advantage that any type of operators, in particular, Schr\"odinger
or Dirac operators can be treated.
We start with  the direct orthogonal sum $L := A \oplus T$
of two symmetric operators $A$ and $T$ with equal deficiency indices
acting in the Hilbert spaces $\gotH$ and $\gotK$, respectively.
From a physical point of view the systems $\{A,\gotH\}$ and
$\{T,\gotK\}$  can be regarded as incomplete internal and external
systems, respectively. The system $\{L,\gotL\}$, $\gotL := \gotH
\oplus \gotK$,  is also an incomplete
quantum system which is completed or closed by choosing a selfadjoint
extension of $L$.
The operator $L$ admits several selfadjoint extensions in $\gotL$. In
particular, there are selfadjoint extensions of the form $L_0 = A_0
\oplus T_0$, where $A_0$ and $T_0$ are selfadjoint extensions of $A$
and $T$ in $\gotH$ and $\gotK$, respectively. Of course, in this case the quantum system $\{L_0,\gotL\}$
decomposes into the closed internal and external system 
$\{A_0,\gotH\}$ and $\{T_0,\gotK\}$, respectively, which
do not interact. There are other selfadjoint extensions of $L$ in $\gotL$ which
are not of this structure and can be regarded as Hamiltonians
of quantum systems which take into account a certain interaction of
the internal and external systems $\{A,\gotH\}$ and $\{T,\gotK\}$. 
In the following we choose a special self-adjoint extension
$\widetilde{L}$ of $L$ introduced in \cite{DHMS00} and 
used in \cite{BMN2}, see also Theorem~\ref{srzthm}, 
which gives the right physical Hamiltonian in applications. 

For example, let the internal system $\{A,\gotH\}$ and external system
$\{T,\gotK\}$ be given by
the minimal second order differential operators
$A=-\tfrac{d^2}{dx^2}+v$ and $T=-\frac{d^2}{dx^2}+V$
in $\gotH=L^2((x_l,x_r))$ and $\gotK=L^2(\dR\backslash(x_l,x_r))$, 
where $(x_l,x_r)$ is a finite interval and $v,V$ are real potentials.
The extension $L_0$ can be chosen to be 
the direct sum of the selfadjoint
extensions of $A$ and $T$ corresponding to Dirichlet boundary
conditions at $x_l$ and $x_r$. According to \cite {BMN2,DHMS00}  the selfadjoint
extension $\widetilde{L}$ coincides in this case
with the usual selfadjoint Schr\"{o}dinger operator
\begin{equation*}
\widetilde L=-\frac{d^2}{dx^2}+\widetilde v,\qquad \widetilde v(x):=\begin{cases}v(x), & x\in (x_l,x_r),\\
V(x), & x \in \dR\backslash (x_l,x_r),\end{cases}
\end{equation*}
in $\gotL=L^2(\dR)$, cf. Section~\ref{coupsturm}.

Let again $A$ and $T$ be symmetric operators with equal deficiency indices in $\sH$ and $\sK$, respectively.
It will be assumed that the deficiency indices of $A$ and $T$ are finite. 
Then the selfadjoint operator $\widetilde L$ is a finite rank perturbation
in resolvent sense of $L_0 = A_0 \oplus T_0$ and therefore
$\{\widetilde L,L_0\}$ is a complete scattering system, i.e., 
the wave operators
\begin{equation*}
W_\pm(\widetilde{L},L_0) := 
\slim_{t\to\pm\infty}e^{it\widetilde{L}}e^{-itL_0}P^{ac}(L_0)
\end{equation*}
exist and map onto the absolutely continuous subspace $\gotH^{ac}(\widetilde L)$ of $\widetilde L$, 
where $P^{ac}(L_0)$ is the orthogonal projection onto
$\gotH^{ac}(L_0)$, cf. \cite{BW}.
The scattering operator 
\begin{equation*}
S:=W_+(\widetilde{L},L_0)^*W_-(\widetilde{L},L_0)
\end{equation*}
regarded as an unitary operator in the absolutely continuous subspace $\gotH^{ac}(L_0)$ is
unitarily equivalent to a multiplication operator induced by a family of unitary matrices 
$\{S(\lambda)\}_{\lambda\in\dR}$ in a spectral representation
of the absolutely continuous part of $L_0$. This multiplication operator $\{S(\lambda)\}_{\lambda\in\dR}$
is called the scattering matrix of the scattering system $\{\widetilde
L,L_0\}$ and is one of the most important objects in
mathematical scattering theory. 
The case that the spectrum $\sigma(A_0)$ is
discrete is of particular importance in physical applications, e.g.,
modeling of quantum transport in semiconductors. In this case the
scattering matrix of $\{\widetilde L,L_0\}$ is given by
\begin{equation*}
S(\lambda)=
I-2i\sqrt{\imag(\tau(\lambda))}\bigl(M(\lambda)+\tau(\lambda)\bigr)^{-1}\sqrt{\imag(\tau(\lambda))},
\end{equation*}
where $M(\cdot)$ and $\tau(\cdot)$ are certain "abstract"
Titchmarsh-Weyl functions corresponding to  
the internal and external systems, respectively, see Corollary
\ref{srzthmcor}. 

The $R$-matrix  $\{R(\gl)\}_{\gl \in \bR}$ of $\{\widetilde L,L_0\}$ 
is defined as the Cayley transform of the scattering matrix $\{S(\gl)\}_{\gl \in \bR}$, i.e.,
\begin{displaymath}
R(\gl) = i(I - S(\gl))(I + S(\gl))^{-1},
\end{displaymath}
and the problem in the $R$-matrix theory is to represent $\{R(\gl)\}_{\gl \in \bR}$
in terms of eigenvalues and eigenfunctions of a suitable
chosen closed internal system $\{\widehat{A},\gotH\}$. By the inverse Cayley transform
this immediately also yields 
a representation of the scattering matrix by the same quantities.

For Schr\"odinger operators the problem is usually solved by choosing 
appropriate selfadjoint boundary conditions at the interface between the internal
and external system, in particular, Neumann boundary conditions. 
We show that in the abstract approach to the $R$-matrix theory the problem can be solved within  
the framework of abstract boundary triplets, which allow to characterize all
selfadjoint extensions of $A$ by abstract boundary
conditions, cf. \cite{DM87,DM91,DM95,GG}. It is one of our main objectives 
to prove that there always exists a family of closed internal systems
$\{A(\gl),\gotH\}_{\gl \in \bR}$ given by abstract boundary conditions connected with
the function $\tau(\cdot)$,
such that the $R$-matrix $\{R(\gl)\}_{\gl \in \bR}$ and the scattering matrix
$\{S(\gl)\}_{\gl \in \bR}$ of $\{\widetilde L,L_0\}$ 
can be expressed with the help of the eigenvalues and eigenfunctions of $A(\gl)$ for
a.e. $\gl \in \bR$, cf. Theorem~\ref{proppo}. This representation requires in addition that the internal
Hamiltonians $A(\gl)$ satisfy $A(\gl) \le A_0$, which is always true if $A_0$ is the Friedrichs
extension of $A$. Moreover, our general representation results
also indicate that even for small energy ranges it is rather unusual that the 
$R$-matrix and the scattering matrix can be represented by
the eigenvalues and eigenfunctions of a single $\gl$-independent
internal Hamiltonian~$\widehat{A}$.

As an application 
again the second order differential operators $A=-\tfrac{d^2}{dx^2}+v$ and $T=-\tfrac{d^2}{dx^2}+V$ 
from above are investigated and particular attention is paid to the case where the
potential $V$ is a real constant. Then
the family $\{A(\lambda)\}_{\lambda\in\dR}$ reduces to 
a single selfadjoint operator, namely, to the Schr\"{o}dinger operator in $L^2((x_l,x_r))$ with
Neumann boundary conditions. In general, however,
this is not the case. Indeed, even in the simple case where $V$ is constant on $(-\infty,x_l)$ and
$(x_r,\infty)$ but the constants are different, a
$\gl$-dependent family of internal Hamiltonians is required for
a certain energy interval to obtain a representation of the  $R$-matrix
and the scattering matrix in terms of eigenfunctions, see Section~\ref{vlvr}.
The condition $A(\gl) \le A_0$ is always
satisfied if $A_0$ is chosen to be 
the Schr\"{o}dinger operator with Dirichlet boundary conditions. Finally, we note that it is
not possible to represent the $R$-matrix and the scattering matrix in terms of
eigenfunctions of an internal Hamiltonian with Dirichlet boundary conditions.

The paper is organized as follows. In Section \ref{btrips} we briefly recall 
some basic facts on boundary triplets and associated Weyl functions corresponding
to symmetric operators in Hilbert spaces. It is the
aim of the simple examples from semiconductor modeling in 
Section~\ref{sturm} to make the reader more familiar with
this efficient tool in extension and spectral theory of symmetric and selfadjoint operators.
Section~\ref{drei} deals with semibounded extensions and
representations of Weyl functions in terms of eigenfunctions of 
selfadjoint extensions of a given symmetric operator. In
Section~\ref{scatsec} we prove general representation
theorems for the scattering matrix and the $R$-matrix of a scattering system which consists of two selfadjoint
extensions of the same symmetric operator. Section~\ref{coupsys} 
is devoted to scattering theory in open quantum systems,
and with the preparations from the previous sections we easily obtain
the abovementioned representation of
the $R$-matrix and scattering matrix of $\{\widetilde L,L_0\}$ in
terms of the eigenfunctions of an energy dependent selfadjoint 
operator family. In the last section the general results are applied to scattering systems consisting
of orthogonal sums of regular and singular ordinary second order differential operators.

\section{Boundary triplets and Weyl functions}\label{btrips}

\subsection{Boundary triplets}

Let $\gotH$ be a separable Hilbert space and let 
$A$ be a densely defined closed symmetric operator with equal
deficiency indices $n_\pm(A)=\dim\ker(A^*\mp i)\leq\infty$ in 
$\gotH$. We use the concept of boundary
triplets for the description of the closed extensions of $A$ in
$\gotH$, see e.g. \cite{DM87,DM91,DM95,GG}.

\begin{defn}
{\em
Let $A$ be a densely defined closed symmetric operator in $\gotH$.
A triplet $\Pi=\{\kH,\gG_0,\gG_1\}$ is called a {\rm boundary triplet} for the adjoint
operator $A^*$ if $\kH$ is a Hilbert space and
$\Gamma_0,\Gamma_1:\  \dom(A^*)\rightarrow\kH$ are linear mappings such that
the abstract Green's identity,
\begin{equation*}
(A^*f,g) - (f,A^*g) = (\gG_1f,\gG_0g) - (\gG_0f,\gG_1g),
\end{equation*}
holds for all $f,g\in\dom(A^*)$ and the mapping
$\gG:=
\bigl(\begin{smallmatrix}
\Gamma_0\\
\Gamma_1
\end{smallmatrix}\bigr):  \dom(A^*) \rightarrow 
\kH\oplus\kH$ is surjective.
}
\end{defn}

We refer to \cite{DM91} and \cite{DM95} for a detailed study of
boundary triplets and recall only some important facts. First of all
a boundary triplet $\Pi=\{\kH,\gG_0,\gG_1\}$ for $A^*$ always exists
since the deficiency indices $n_\pm(A)$ of $A$ are assumed to be equal. In this case 
$n_\pm(A) = \dim\kH$ holds. We also note that a boundary triplet for
$A^*$ is not unique.

In order to describe the set of closed extensions $\widehat A\subseteq A^*$ of $A$ with
the help of a boundary triplet $\Pi=\{\kH,\Gamma_0,\Gamma_1\}$ for
$A^*$ we introduce the set $\widetilde\kC(\kH)$ of closed
linear relations in $\kH$, that is, the set of closed linear
subspaces of $\kH \oplus \kH$. If $\gT$ is a closed linear operator in
$\kH$, then $\gT$ will be  
identified with its graph $\cG(\gT)$,
\begin{equation*}
\Theta\,\,\widetilde=\,\,\cG(\gT) = \left\{\begin{pmatrix} h\\ \gT h\end{pmatrix}: h \in
\dom(\gT)\right\}.
\end{equation*}
Therefore, the set of closed
linear operators in $\kH$ is a subset of
$\widetilde\kC(\kH)$. Note that $\Theta\in\widetilde\cC(\cH)$ is the graph of an operator
if and only if the multivalued part $\mul(\Theta):=\bigl\{h^\prime\in\cH:
\bigl(\begin{smallmatrix}0\\ h^\prime\end{smallmatrix}\bigr)\in\Theta\bigr\}$ is trivial. 
The resolvent set $\rho(\gT)$ and the point, continuous and residual spectrum 
$\sigma_p(\Theta)$, $\sigma_c(\Theta)$ and $\sigma_r(\gT)$ of a closed linear relation $\Theta$
are defined in a similar way as for closed linear operators, cf. \cite{DS87}. Recall that the adjoint
relation $\Theta^*\in\widetilde\kC(\kH)$ of a linear relation
$\Theta$ in $\kH$ is defined as
\be\la{thetastar}
\Theta^*:= \left\{
\begin{pmatrix}
k\\
k^\prime
\end{pmatrix}
: (h^\prime,k)=(h,k^\prime)\,\,\text{for all}\,
\begin{pmatrix}
h\\
h^\prime
\end{pmatrix} \in\Theta\right\}
\end{equation}
and $\Theta$ is said to be {\it symmetric} ({\it selfadjoint}) if
$\Theta\subseteq\Theta^*$ (resp. $\Theta=\Theta^*$). We note that
definition \eqref{thetastar} extends the usual definition of the adjoint operator.
Let now $\Theta$ be a selfadjoint relation in $\kH$ and 
let $P_\op$
be the orthogonal projection in $\kH$ onto
$\cH_\op:=(\mul(\Theta))^\bot=\overline{\dom(\Theta)}$. Then
\begin{equation*}
\Theta_\op=\left\{\begin{pmatrix}x\\ P_\op x^\prime\end{pmatrix}:\begin{pmatrix} x\\x^\prime
\end{pmatrix}\in\Theta\right\}
\end{equation*}
is a selfadjoint (possibly unbounded) operator in the Hilbert space 
$\cH_\op$ and $\Theta$ can be written as the direct orthogonal sum of $\Theta_\op$
and a "pure" relation $\Theta_\infty$ in the Hilbert space
$\cH_\infty:=(1-P_\op)\cH=\mul\Theta$,
\begin{equation}\label{thetadeco}
\Theta=\Theta_\op\oplus\Theta_\infty,\qquad\Theta_\infty:=\left\{\begin{pmatrix}0\\ x^\prime\end{pmatrix}
:x^\prime\in\mul\Theta\right\}\in\widetilde \cC(\cH_\infty).
\end{equation}
With a  boundary triplet  $\Pi=\{\kH,\gG_0,\gG_1\}$ for $A^*$ one  associates
two selfadjoint extensions of $A$ defined by
\begin{equation}\label{2.2}
A_0:=A^*\!\upharpoonright\ker(\gG_0)
\quad \text{and}\quad
A_1:=A^*\!\upharpoonright\ker(\gG_1).
\end{equation}
A description of all proper (symmetric, selfadjoint) extensions
of $A$ is given in the next proposition. 
\begin{prop}\label{propo}
Let $A$ be a densely defined closed symmetric operator in $\gotH$ with equal deficiency indices and
let
$\Pi=\{\kH,\gG_0,\gG_1\}$ be a boundary triplet for  $A^*.$  Then the mapping
\begin{equation}\label{bij}
\Theta\mapsto A_\Theta:= A^*\upharpoonright \Gamma^{(-1)}\Theta=
A^*\upharpoonright \bigl\{f\in\dom(A^*): \  (\Gamma_0
f,\Gamma_1 f)^\top\in\Theta\bigr\}
\end{equation}
establishes  a bijective correspondence between the set
$\widetilde\kC(\kH)$ and the set of closed extensions $A_\Theta\subseteq A^*$ of $A$.
Furthermore
\begin{equation*}
(A_\Theta)^*=  A_{\Theta^*}
\end{equation*}
holds for any $\Theta\in\widetilde\kC(\kH)$. The extension $A_\Theta$ in \eqref{bij}
is symmetric (selfadjoint, dissipative,
maximal dissipative) if and only if $\Theta$ is symmetric (selfadjoint, dissipative,
maximal dissipative).
\end{prop}
It is worth to note that the selfadjoint operator $A_0=A^*\upharpoonright\ker(\Gamma_0)$ in \eqref{2.2} 
corresponds to the "pure" relation
$\Theta_\infty=\bigl\{\bigl(\begin{smallmatrix} 0\\ h\end{smallmatrix}\bigr):h\in\kH \bigr\}$. Moreover,
if $\Theta$ is an operator, then \eqref{bij} can also be written in
the form
\begin{equation}\label{bij2}
A_\Theta= A^*\upharpoonright\ker\bigl(\Gamma_1-\Theta\Gamma_0\bigr),
\end{equation}
so that, in particular $A_1$ in \eqref{2.2} corresponds to $\Theta=0\in[\cH]$. Here and in the following
$[\cH]$ stands for the space of bounded everywhere defined linear operators in $\cH$. 
We note that if the product $\Theta\Gamma_0$ in \eqref{bij2}
is interpreted in the sense of relations, then \eqref{bij2} is even true for parameters $\Theta$ with
$\mul(\Theta)\not=\{0\}$.

Later we shall often be concerned with closed simple
symmetric operators. Recall that a closed symmetric operator $A$ is said
to be {\it simple} if there is no nontrivial subspace which reduces
$A$ to a selfadjoint operator. By \cite{K49} this is equivalent to
\begin{equation*}
\gotH=\clo\spa\bigl\{\ker(A^*-\gl): \gl\in\bC\backslash\bR\bigr\},
\end{equation*}
where $\clospa \{\cdot\}$ denotes the closed linear span of a set. Note that a simple symmetric
operator has no eigenvalues.

\subsection{Weyl functions and resolvents of extensions}\label{weylreso}

Let again $A$ be a densely defined closed
symmetric operator in $\gotH$ with equal deficiency indices. 
A point $\lambda\in\dC$ is of {\it regular type} if $\ker(A-\lambda)=\{0\}$ and 
the range $\ran(A- \gl)$ is closed. We
denote the {\it defect subspace} of $A$ at the points $\gl \in
\bC$ of regular type by $\kN_\gl=\ker(A^*-\gl)$. The space of bounded everywhere defined linear operators 
mapping the Hilbert space $\cH$ into $\sH$ will be denoted by $[\cH,\sH]$. The following definition was given in 
\cite{DM87,DM91}.

\begin{defn}\label{Weylfunc}
{\em
Let $A$ be  a densely defined closed
symmetric operator in $\gotH$,
let $\Pi=\{\kH,\gG_0,\gG_1\}$ be a boundary triplet for $A^*$ and
let $A_0=A^*\!\upharpoonright\ker(\gG_0)$. The operator-valued
functions
$\gamma(\cdot):\rho(A_0)\rightarrow  [\kH,\gotH]$ and  $M(\cdot):
\rho(A_0)\rightarrow  [\kH]$ defined by
\begin{equation}\la{2.3A}
\gamma(\gl):=\bigl(\Gamma_0\!\upharpoonright\kN_\gl\bigr)^{-1} \qquad\text{and}\qquad
M(\gl):=\Gamma_1\gamma(\gl), \quad \gl\in\rho(A_0),
\end{equation}
are called the {\rm $\gamma$-field} and the {\rm Weyl function}, respectively,
corresponding to the boundary triplet $\Pi$.
}
\end{defn}

It follows from the identity  $\dom(A^*)=\ker(\Gamma_0)\dot +\kN_\gl$,
$\lambda\in\rho(A_0)$, where as above $A_0=A^*\!\upharpoonright\ker(\gG_0)$,
that the $\gamma$-field  $\gamma(\cdot)$ in \eqref{2.3A} is well defined.
It is easily seen that both  $\gamma(\cdot)$ and $M(\cdot)$ are
holomorphic on $\rho(A_0)$, and the relations
\begin{equation*}
\gamma(\lambda)=\bigl(1+(\gl-\mu)(A_0-\gl)^{-1}\bigr)\gamma(\mu),
\qquad \gl,\mu\in\rho(A_0),
\end{equation*}
and
\begin{equation}\la{mlambda}
M(\gl)-M(\mu)^*=(\gl-\overline\mu)\gamma(\mu)^*\gamma(\gl),
\qquad \gl,\mu\in\rho(A_0),
\end{equation}
are valid (see \cite{DM91}).
The identity \eqref{mlambda} yields that $M(\cdot)$ is a  {\it
Nevanlinna function}, that is, $M(\cdot)$ is holomorphic on $\bC\backslash\bR$, 
$M(\gl)=M(\overline\gl)^*$ for
all $\gl\in\bC\backslash\bR$ and $\imag(M(\gl))$ is a nonnegative operator
for all $\gl$ in the upper half plane $\bC_+=\{\lambda\in\bC:\imag (\lambda)>0\}$.
Moreover, it follows from \eqref{mlambda}
that $0\in \rho(\imag(M(\gl)))$ holds for all $\lambda\in\dC\backslash\dR$. 

The following well-known theorem shows how the spectral properties
of the closed extensions $A_\Theta$ of $A$ can be described with the help
of the Weyl function, cf. \cite{DM91,DM95}.

\begin{thm}\label{resthm}
Let $A$ be a densely defined closed symmetric operator in $\gotH$ and let $\{\cH,\Gamma_0,\Gamma_1\}$ be
a boundary triplet for $A^*$ with $\gamma$-field $\gamma$ and Weyl function $M$. Let 
$A_0=A^*\upharpoonright\ker(\Gamma_0)$
and let $A_\Theta\subseteq A^*$ be a closed extension
corresponding to some $\Theta\in\widetilde\kC(\kH)$ via \eqref{bij}-\eqref{bij2}. Then
a point $\gl\in\rho(A_0)$ belongs to the resolvent set 
$\rho(A_\Theta)$ if and only if $0\in\rho(\Theta-M(\gl))$ and the
formula 
\begin{equation}\la{2.8}
(A_\gT - \gl)^{-1} = (A_0 - \gl)^{-1} + \gga(\gl)\bigl(\gT - M(\gl)\bigr)^{-1}\gga(\overline{\gl})^*
\end{equation}
holds for all $\gl \in \rho (A_0) \cap \rho (A_\gT)$. Moreover, $\gl$
belongs to the point spectrum $\gs_p(A_\gT)$, to the continuous
spectrum $\gs_c(A_\gT)$ or to the residual spectrum $\gs_r(A_\gT)$
if and only if $0\in\sigma_i(\Theta-M(\gl))$, $i=p,c,r$, respectively.
\end{thm}

\subsection{Regular and singular Sturm-Liouville operators}\label{sturm}

We are going to illustrate the notions of boundary
triplets, Weyl functions and $\gga$-fields with some well-known simple examples.

\subsubsection{Finite intervals}\la{2.3.1}

Let us first consider a Schr\"odinger operator on the bounded interval
$(x_l,x_r) \subset \bR$. The minimal operator $A$ in $\gotH = L^2((x_l,x_r))$ is defined by
\begin{equation}\la{inner1}
\begin{split}
(Af)(x) &:=
-\frac{1}{2}\frac{d}{dx}\frac{1}{m(x)}\frac{d}{dx}f(x) + v(x)f(x),\\
\dom(A) & :=  \left\{f \in\gotH: \ba{l}
f, \frac{1}{m}f' \in W^{1,2}((x_l,x_r)) \\
f(x_l) = f(x_r)= 0\\
\left(\frac{1}{m}f'\right)(x_l) = \left(\frac{1}{m}f'\right)(x_r) = 0
\ea
\right\},
\end{split}
\end{equation}
where it is assumed that the effective mass $m$ satisfies $m > 0$ and $m,
\tfrac{1}{m} \in L^\infty((x_l,x_r))$, and that $v \in
L^\infty((x_l,x_r))$ is a real function.
It is well known that $A$ is a densely defined closed simple symmetric operator in
$\gotH$ with deficiency indices $n_+(A) = n_-(A) = 2$.  The adjoint
operator $A^*$ is given by
\bed
\begin{split}
(A^*f)(x) & =  -\frac{1}{2}\frac{d}{dx}\frac{1}{m(x)}\frac{d}{dx}f(x) + v(x)f(x),\\
\dom(A^*) & =  \left\{f \in \gotH: f,\tfrac{1}{m}f' \in
W^{1,2}((x_l,x_r))\right\}.
\end{split}
\eed
It is straightforward to verify that $\Pi_A=\{\bC^2,\gG_0,\gG_1\}$, where
\begin{equation*}
\gG_0f := \left(
\begin{array}{c}
f(x_l)\\
f(x_r)
\end{array}
\right)
\quad \mbox{and} \quad
\gG_1f := \frac{1}{2}\left(
\begin{array}{c}
\left(\frac{1}{m}f'\right)(x_l)\\
-\left(\frac{1}{m}f'\right)(x_r)
\end{array}
\right),
\end{equation*}
$f\in \dom(A^*)$, is a boundary triplet for $A^*$.
Note, that the selfadjoint extension
$A_0 := A^*\!\upharpoonright\ker(\gG_0)$ corresponds to Dirichlet
boundary conditions, that is,
\begin{equation}\la{2.14}
\dom(A_0) = \left\{f \in \gotH: f,\tfrac{1}{m}f' \in
W^{1,2}((x_l,x_r)), \,f(x_l) = f(x_r) = 0 \right\}.
\end{equation}
The selfadjoint extension $A_1$ corresponds to Neumann boundary conditions, i.e., 
\begin{equation}\label{2.33b}
\dom(A_1) =
\left\{f \in \gotH: \ba{l} f,\tfrac{1}{m}f' \in
W^{1,2}((x_l,x_r)),\\ (\tfrac{1}{m}f')(x_l) = (\tfrac{1}{m}f')(x_r) = 0\ea
\right\}.
\end{equation}
Let $\varphi_\gl$ and $\psi_\gl$, $\lambda\in\dC$, be the fundamental solutions 
of the homogeneous differential equation $-\frac{1}{2}\frac{d}{dx}\frac{1}{m}\frac{d}{dx}u + v\,u =
\gl u$ satisfying the boundary conditions
\begin{equation*}
\varphi_\gl(x_l) = 1,\,\,\, (\tfrac{1}{2m}\varphi_\lambda^\prime)(x_l) = 0\quad\text{and}\quad
\psi_\gl(x_l) = 0,\,\,\,(\tfrac{1}{2m}\psi'_\gl)(x_l) = 1.
\end{equation*}
Note that $\varphi_\gl$ and $\psi_\gl$ belong to $L^2((x_l,x_r))$ since $(x_l,x_r)$ is a finite interval. 
A straightforward computation shows 
\begin{equation*}
\begin{split}
\bigl((A_0 -\gl)^{-1}f\bigr)(x) &=\varphi_\gl(x)\int^x_{x_l} \psi_\gl(t)f(t) \,dt +
\psi_\gl(x)\int^{x_r}_x \varphi_\gl(t)f(t) \,dt  \\
&\qquad\qquad\qquad -\frac{\varphi_\gl(x_r)}{\psi_\gl(x_r)}\,\psi_\gl(x)\int^{x_r}_{x_l}\psi_\gl(t)f(t) \, dt
\end{split}
\end{equation*}
for $x \in (x_l,x_r)$, $f \in L^2((x_l,x_r))$ and all $\gl \in \rho(A_0)$.
In order to calculate the $\gamma$-field and Weyl function corresponding to
$\Pi_A=\{\bC^2,\gG_0,\gG_1\}$ note that every element $f_\gl \in \kN_\gl = \ker(A^* - \gl)$ admits the
representation
\begin{equation*}
f_\gl(x) = \xi_0\varphi_\gl(x) + \xi_1\psi_\gl(x), \quad x \in (x_l,x_r),
\quad \gl \in \bC, \quad \xi_0,\xi_1 \in \bC,
\end{equation*}
where the coefficients $\xi_0,\xi_1$ are uniquely determined.
The relation 
\bed
\gG_0f_\gl =
\begin{pmatrix}
1 & 0\\
\varphi_\gl(x_r) & \psi_\gl(x_r)
\end{pmatrix}
\begin{pmatrix}
\xi_0\\
\xi_1
\end{pmatrix}
\eed
yields
\begin{equation*}
\frac{1}{\psi_\gl(x_r)}
\begin{pmatrix}
\psi_\gl(x_r) & 0\\
-\varphi_\gl(x_r) & 1
\end{pmatrix}
\gG_0f_\gl = 
\begin{pmatrix}
\xi_0\\
\xi_1
\end{pmatrix}
\end{equation*}
for $\psi_\lambda(x_r)\not=0$ (that is $\lambda\not\in\sigma(A_0)$)
and it follows that the $\gga$-field is given by
\begin{equation*}
\begin{split}
\gamma(\lambda):\dC^2&\rightarrow L^2((x_l,x_r)),\\
\begin{pmatrix}
\xi_0\\
\xi_1
\end{pmatrix}
&\mapsto
\frac{1}{\psi_\gl(x_r)}
\bigl(
(\varphi_\gl(\cdot)\psi_\gl(x_r) - \psi_\gl(\cdot)\varphi_\gl(x_r))\xi_0+  \psi_\gl(\cdot)\xi_1
\bigr).
\end{split}
\end{equation*}
We remark that the adjoint operator admits the representation
\bed
\gga(\gl)^*f = 
\frac{1}{\overline{\psi_\gl(x_r)}}
\begin{pmatrix}
\int^{x_r}_{x_l}\big(\overline{\varphi_\gl(y)}\;\overline{\psi_\gl(x_r)} 
- \overline{\psi_\gl(y)}\;\overline{\varphi_\gl(x_r)}\big)f(y)\, dy\\
\int^{x_r}_{x_l} \overline{\psi_\gl(y)}f(y)\,dy
\end{pmatrix},
\eed
$f \in L^2((x_l,x_r))$. The Weyl function $M(\lambda)=\Gamma_1\gamma(\lambda)$,
$\lambda\in\rho(A_0)$, then becomes
\bed
M(\gl) = \frac{1}{\psi_\gl(x_r)} 
\begin{pmatrix}
-\varphi_\gl(x_r) & 1\\
1 & -(\frac{1}{2m}\psi'_\gl)(x_r)
\end{pmatrix}.
\eed
All selfadjoint extension of $A$ can now be described with the help of selfadjoint relations
$\gT = \gT^*$ in $\dC^2$ via \eqref{bij}-\eqref{bij2} and their resolvents can be expressed in terms
of the resolvent of $A_0$, the Weyl function $M(\cdot)$ and the
$\gamma$-field $\gamma(\cdot)$, 
cf. Theorem~\ref{resthm}.
We leave the general case to the reader and note only that if $\Theta$ is a selfadjoint matrix of the form
\bed
\gT = 
\begin{pmatrix}
\gk_l & 0\\
0 & \gk_r
\end{pmatrix},
\quad \gk_l,\gk_r \in \bR,
\eed
then
\bed
\dom(A_\gT) = \left\{ f \in \dom(A^*): 
\begin{matrix}
(\frac{1}{2m}f')(x_l) = \gk_lf(x_l)\\
(\frac{1}{2m}f')(x_r) = -\gk_rf(x_r)
\end{matrix}
\right\}
\eed
and 
\bead
\lefteqn{
\bigl(\gT - M(\gl)\bigr)^{-1} = }\\
& &
\frac{1}{\psi_\gl(x_r) \det(\gT - M(\gl))}
\begin{pmatrix}
\gk_r\psi_\gl(x_r) + (\frac{1}{2m}\psi'_\gl)(x_r) & 1\\
1 & \gk_l\psi_\gl(x_r) + \varphi_\gl(x_r)
\end{pmatrix}.
\eead
Obviously the case $\kappa_l=\kappa_r=0$ leads to the Neumann operator $A_1$.

\subsubsection{Infinite intervals}\la{2.3.2}

Next we consider a singular problem on the infinite interval
$(-\infty,x_l)$ in the Hilbert space $\gotK_l = L^2((-\infty,x_l))$. The minimal operator is defined by
\begin{equation*}
\begin{split}
(T_lg_l)(x) &:=
-\frac{1}{2}\frac{d}{dx}\frac{1}{m_l(x)}\frac{d}{dx}g_l(x) + v_l(x)g_l(x),\\
\dom(T_l) & :=  \left\{g_l \in\gotK_l: \ba{l}
g_l, \frac{1}{m_l}g_l' \in W^{1,2}((-\infty,x_l)) \\
g_l(x_l) = \bigl(\tfrac{1}{m_l}g_l'\bigr)(x_l) = 0
\ea
\right\},
\end{split}
\end{equation*}
where $m_l > 0$, $m_l,\tfrac{1}{m_l}\in L^\infty((-\infty,x_l))$ and $v_l \in
L^\infty((-\infty,x_l))$ is real. Then $T_l$ is a densely defined closed simple symmetric operator
with deficiency indices $n_-(T_l) = n_+(T_l) = 1$, 
see e.g. \cite{Wei1} and \cite{G72} for the fact that $T_l$ is simple. 
The adjoint
operator $T^*$ is given by
\begin{equation*}
\begin{split}
(T_l^*g_l)(x) &=
-\frac{1}{2}\frac{d}{dx}\frac{1}{m_l(x)}\frac{d}{dx}g_l(x) + v_l(x)g_l(x),\\
\dom(T_l^*) & =  \left\{g_l \in\gotK_l: 
g_l, \tfrac{1}{m_l}g_l' \in W^{1,2}((-\infty,x_l)) 
\right\}.
\end{split}
\end{equation*}
One easily verifies that $\gP_{T_l} = \{\bC,\gY^l_0,\gY^l_1\}$,
\bed
\gY^l_0g_l := g_l(x_l) \quad \mbox{and} \quad \gY^l_1g_l :=
-\left(\frac{1}{2m_l}g_l'\right)(x_l), \quad g_l \in \dom(T^*_l),
\eed
is a boundary triplet for $T_l^*$. Let $\varphi_{\lambda,l}$ and $\psi_{\lambda,l}$ be the 
fundamental solutions
of the equation $-\frac{1}{2}\frac{d}{dx}\frac{1}{m_l}\frac{d}{dx}u + v_lu = \gl u$ satisfying the
boundary conditions
\begin{equation*}
\varphi_{\gl,l}(x_l) = 1,\,\,\,
\left(\tfrac{1}{2m_l}\varphi_{\gl,l}^\prime \right)(x_l) = 0
\quad \text{and} \quad 
\psi_{\gl,l}(x_l) = 0,\,\,\, 
\left(\tfrac{1}{2m_l}\psi_{\gl,l}^\prime\right)(x_l) = 1. 
\end{equation*}
Then there exists a scalar function $\gotm_l$ such that for each $\gl \in \bC
\setminus \bR$ the function 
\bed
x\mapsto g_{\gl,l}(x) := \varphi_{\gl,l}(x) - \gotm_l(\gl)\psi_{\gl,l}(x)
\eed
belongs to $L^2((-\infty,x_l))$, cf. \cite{Wei1}. 
The function $\gotm_l$ is usually called the Titchmarsh-Weyl function
or Titchmarsh-Weyl coefficient and in our setting $\gotm_l$
coincides with the Weyl function of the boundary triplet $\gP_{T_l} = \{\bC,\gY^l_0,\gY^l_1\}$, since
\bed
\gY^l_1 g_{\gl,l} = \gotm_l(\gl)\gY^l_0 g_{\gl,l},\quad 
g_{\gl,l} \in \kN_{\gl,l} := \ker(T^*_l - \gl), \quad \gl \in \bC \setminus \bR.
\eed

An analogous example is the Schr\"odinger operator on the infinite interval
$(x_r,\infty)$ in $\gotK_r = L^2((x_r,\infty))$ defined by
\begin{equation*}
\begin{split}
(T_rg_r)(x) &:=
-\frac{1}{2}\frac{d}{dx}\frac{1}{m_r(x)}\frac{d}{dx}g_r(x) + v_r(x)g_r(x),\\
\dom(T_r) & :=  \left\{g_r \in\gotK_r: \ba{l}
g_r, \frac{1}{m_r}g_r' \in W^{1,2}((x_r,\infty)) \\
g_r(x_r) = \bigl(\frac{1}{m_r}g_r'\bigr)(x_r) = 0
\ea
\right\},
\end{split}
\end{equation*}
where $m_r > 0$, $m_r,\tfrac{1}{m_r}\in L^\infty((x_r,\infty))$ and $v_r \in
L^\infty((x_r,\infty))$ is real. The adjoint
operator $T_r^*$ is 
\begin{equation*}
\begin{split}
(T_r^*g_r)(x) &=
-\frac{1}{2}\frac{d}{dx}\frac{1}{m_r(x)}\frac{d}{dx}g_r(x) + v_r(x)g_r(x),\\
\dom(T_r^*) & =  \left\{g_r \in\gotK_r: 
g_r, \tfrac{1}{m_r}g_r' \in W^{1,2}((x_r,\infty)) 
\right\}
\end{split}
\end{equation*}
and $\gP_{T_r} = \{\bC,\gY^r_0,\gY^r_1\}$,
\bed
\gY^r_0g_r := g_r(x_r) \quad \mbox{and} \quad \gY^r_1g_r :=
\left(\frac{1}{2m_r}g_r'\right)(x_r), \quad g_r \in \dom(T^*_r),
\eed
is a boundary triplet for $T_r^*$. Let $\varphi_{\lambda,r}$ and $\psi_{\lambda,r}$ be the 
fundamental solutions
of the equation $-\frac{1}{2}\frac{d}{dx}\frac{1}{m_r}\frac{d}{dx}u + v_ru = \gl u$ satisfying the
boundary conditions
\begin{equation*}
\varphi_{\gl,r}(x_r) = 1,\,\,\,
\left(\tfrac{1}{2m_r}\varphi_{\gl,r}^\prime \right)(x_r) = 0
\quad \text{and} \quad 
\psi_{\gl,r}(x_r) = 0,\,\,\, 
\left(\tfrac{1}{2m_r}\psi_{\gl,r}^\prime\right)(x_r) = 1. 
\end{equation*}
Then there exists a scalar function $\gotm_r$ such that for each $\gl \in \bC
\setminus \bR$ the function 
\bed
x\mapsto g_{\gl,r}(x) := \varphi_{\gl,r}(x) +\gotm_r(\gl)\psi_{\gl,r}(x)
\eed
belongs to $L^2((x_r,\infty))$. As above $\gotm_r$ coincides with the Weyl function
of the boundary triplet $\gP_{T_r} := \{\bC,\gY^r_0,\gY^r_1\}$.

For our purposes it is useful to consider the direct sum of the two operators $T_l$ and
$T_r$. To this end we introduce the Hilbert space 
\begin{equation*}
\gotK := L^2((-\infty,x_l)\cup(x_r,\infty)) \widetilde = \gotK_l \oplus \gotK_r.
\end{equation*}
An element $g \in \gotK$
will be written in the form $g = g_l \oplus g_r$, where $g_l\in
L^2((-\infty,x_l))$ and $g_r\in L^2((x_r,\infty))$.
The operator $T = T_l \oplus T_r$ in $\gotK$ is defined by
\bed
\begin{split}
(Tg)(x) = &
\left(
\ba{cc}
-\frac{1}{2}\frac{d}{dx}\frac{1}{m_l(x)}\frac{d}{dx}g_l(x) + v_l g_l(x) & 0\\
0 & -\frac{1}{2}\frac{d}{dx}\frac{1}{m_r(x)}\frac{d}{dx}g_r(x) + v_r g_r(x)
\ea
\right),\\
\dom(T) & = \dom(T_l) \oplus \dom(T_r),
\end{split}
\eed
and $T$ is a densely defined closed simple symmetric operator in $\gotK$ with deficiency indices
$n_+(T)=n_-(T)=2$. The adjoint operator $T^*$ is given by
\bed
\begin{split}
(T^*g)(x) = &
\left(
\ba{cc}
-\frac{1}{2}\frac{d}{dx}\frac{1}{m_l(x)}\frac{d}{dx}g_l(x) + v_l g_l(x) & 0\\
0 & -\frac{1}{2}\frac{d}{dx}\frac{1}{m_r(x)}\frac{d}{dx}g_r(x) + v_r g_r(x)
\ea
\right),\\
\dom(T^*) & = \dom(T^*_l) \oplus \dom(T^*_r).
\end{split}
\eed
One easily checks that $\Pi_T =\{\bC^2,\gY_0,\gY_1\}$, $\gY_0 :=
(\gY^l_0,\gY^r_0)^\top$, $\gY_1 := (\gY^l_1 ,\gY^r_1)^\top$, that is,
\begin{equation*}
\gY_0 g = \left(
\ba{c}
g_l(x_l)\\
g_r(x_r)
\ea
\right)
\quad \mbox{and} \quad
\gY_1 g = \frac{1}{2}\left(
\ba{c}
-\bigl(\frac{1}{m_l}g'_l\bigr)(x_l)\\
\bigl(\frac{1}{m_r}g'_r\bigr)(x_r)
\ea
\right),
\end{equation*}
$g\in \dom(T^*)$, is a boundary triplet for $T^*$. Note that $T_0 =
T^*\!\upharpoonright\ker(\gY_0)$ is the restriction of $T^*$ to the domain
\begin{equation*}
\dom(T_0) = \bigl\{g\in\dom(T^*):
g_l(x_l) = g_r(x_r) = 0\bigr\},
\end{equation*}
that is, $T_0$ corresponds to Dirichlet boundary conditions at $x_l$ and $x_r$. The Weyl
function $\gt(\cdot)$ corresponding to the boundary triplet $\gP_T=\{\bC^2,\gY_0,\gY_1\}$ is given by 
\begin{equation*}
\gl \mapsto \tau(\gl)=
\begin{pmatrix}
\gotm_l(\gl) & 0\\
0 & \gotm_r(\gl) \end{pmatrix} ,
\qquad\lambda\in\rho(T_0).
\end{equation*}

\section{Semibounded extensions and expansions in eigenfunctions}\label{drei}

Let $A$ be a densely defined closed symmetric operator in the separable Hilbert space $\gotH$ and 
let $\{\cH,\Gamma_0,\Gamma_1\}$ be
a boundary triplet for $A^*$ with $\gamma$-field $\gamma(\cdot)$ and Weyl function $M(\cdot)$. 
Fix some $\Theta=\Theta^*\in\widetilde\cC(\cH)$ and let $A_\Theta\subseteq A^*$ be the corresponding 
selfadjoint extension via \eqref{bij}.

In the next proposition it will be assumed that 
$A_0=A^*\upharpoonright\ker(\Gamma_0)$ and $A_\Theta$ (and hence also the symmetric operator $A$)
are semi-bounded from below. Note that if $A$ has finite defect it is sufficient for this to assume that
$A$ is semibounded, cf. Corollary~\ref{corex}.

\begin{prop}\label{II.4}
Let $A$ be a densely defined closed symmetric operator in $\gotH$ and let 
$\{\kH,\gG_0,\gG_1\}$ be a boundary triplet for $A^*$ with $\gamma$-field $\gamma(\cdot)$
and Weyl function $M(\cdot)$. Let $A_\Theta$ be a selfadjoint extension of $A$ corresponding to
$\Theta=\Theta^*\in\widetilde\cC(\cH)$ and assume that
$A_0=A^*\upharpoonright\ker(\Gamma_0)$ and $A_\Theta$ are semibounded from below.  
Then $A_\gT \le A_0$ holds if and only if
\begin{equation}\la{2.17a}
\ran\bigl(\gga(\gl)\bigl(\gT - M(\gl)\bigr)^{-1}\bigr) \subseteq \dom\bigl(\sqrt{A_\gT - \gl}\,\bigr)
\end{equation}
is  satisfied for all $\gl < \min\{\inf\gs(A_0),\inf\gs(A_\gT)\}$ .
\end{prop}
\begin{proof}
Let $A_\gT \le A_0$. From \eqref{2.8} we get
\bed
(A_\gT - \gl)^{-1} - (A_0 - \gl)^{-1} = \gga(\gl)\bigl(\gT -
M(\gl)\bigr)^{-1}\gga(\gl)^* \ge 0
\eed
for $\gl <\min\{\inf\gs(A_0),\inf\gs(A_\gT)\}$ which yields
\bed
\bigl(\gT - M(\gl)\bigr)^{-1} \ge 0.
\eed
By \cite[Corollary 7-2]{Fu1} there is a contraction $Y$ acting from $\gotH$ into $\kH$ such that
\bed
\bigl(\gT - M(\gl)\bigr)^{-1/2}\gga(\gl)^* = Y(A_\gT - \gl)^{-1/2}.
\eed
Since $\lambda\in\dR$ the adjoint has the form 
\bed
\gga(\gl)\bigl(\gT - M(\gl)\bigr)^{-1/2} = (A_\gT - \gl)^{-1/2}Y^*,
\eed
so that
\bed
\ran\bigl(\gga(\gl)\bigl(\gT - M(\gl)\bigr)^{-1/2}\bigr) \subseteq \dom\bigl(\sqrt{A_\gT - \gl}\,\bigr).
\eed
Therefore
\bead
\lefteqn{
\ran\bigl(\gga(\gl)\bigl(\gT - M(\gl)\bigr)^{-1}\bigr) \subseteq}\\
& &  
\ran\bigl(\gga(\gl)\bigl(\gT - M(\gl)\bigr)^{-1/2}\bigr) \subseteq \dom\bigl(\sqrt{A_\gT - \gl}\,\bigr)
\nonumber 
\eead
and \eqref{2.17a} is proved. 

Conversely, let us assume that condition \eqref{2.17a} is satisfied. 
Then for each $\gl < \min\{\inf\sigma(A_0),\inf\sigma(A_\gT)\}$ the operator
\begin{equation}\label{ftheta}
F^*_\gT(\gl) := \sqrt{A_\gT - \gl}\,\gamma(\lambda)\bigl(\Theta-M(\lambda)\bigr)^{-1}
\end{equation}
is well defined on $\cH$ and closed, and hence bounded. Besides $F^*_\gT(\gl)$ we introduce the
densely defined operator 
\begin{equation}\label{fthe}
\begin{split}
F_\gT(\gl)  &=  \gG_0(A_\gT - \gl)^{-1/2}, \\
\dom(F_\gT(\gl)) & =  \bigl\{f \in \gotH: (A_\gT - \gl)^{-1/2}f \in
\dom(A^*)\bigr\}
\end{split}
\end{equation}
for $\lambda< \inf\sigma(A_\gT)$. 

It follows from \eqref{2.8}, $A_0=A^*\upharpoonright\ker(\Gamma_0)$
and $\gG_0\gga(\gl) = I_\kH$ that
\begin{equation}\label{asd}
\gG_0(A_\gT - \gl)^{-1}=\bigl(\Theta - M(\gl)\bigr)^{-1}\gga(\overline{\gl})^*
\end{equation}
holds for all $\lambda\in \rho(A_0)\cap\rho(A_\Theta)$. 
Thus for $\gl < \min\{\inf\sigma(A_0),\inf\sigma(A_\gT)\}$
\eqref{ftheta} becomes
\begin{equation*}
F^*_\gT(\gl) = \sqrt{A_\gT - \gl}\,\bigl(\gG_0(A_\gT - \gl)^{-1} \bigr)^*
\end{equation*}
and together with \eqref{fthe} we conclude
\begin{equation*}
F_\Theta(\lambda)=\Gamma_0(A_\Theta-\lambda)^{-1/2}\subseteq
\Bigl(\sqrt{A_\gT - \gl}\,\bigl(\gG_0(A_\gT - \gl)^{-1} \bigr)^*\Bigr)^*=\bigl(F^*_\gT(\gl)\bigr)^*.
\end{equation*}
This implies that $F_\gT(\gl)$ admits a bounded everywhere defined extension
$\overline{F}_\gT(\gl)$ for $\gl <
\min\{\inf\gs(A_0),\inf\gs(A_\gT)\}$ such that $F_\gT(\gl)^* =
\overline{F}_\gT(\gl)^* = F^*_\gT(\gl)$. From \eqref{asd} and $M(\overline\lambda)=M(\lambda)^*$ we find
\begin{equation*}
\gG_0\bigl(\gG_0(A_\gT - \overline\gl)^{-1}\bigr)^*=\bigl(\Theta - M(\gl)\bigr)^{-1},\quad 
\lambda\in \rho(A_0)\cap\rho(A_\Theta),
\end{equation*}
so that for $\gl < \min\{\inf\gs(A_0),\inf\gs(A_\gT)\}$
\begin{equation*}
\begin{split}
\bigl(\Theta - M(\gl)\bigr)^{-1}
&= \gG_0(A_\gT - \gl)^{-1/2} \sqrt{A_\gT -
  \gl}\, \bigl(\gG_0(A_\gT - \gl)^{-1}\bigr)^*  \\
&= \overline{F}_\gT(\gl) \overline{F}_\gT(\gl)^* \ge 0.
\end{split}
\end{equation*}
Using \eqref{2.8} we find
\bed
(A_\gT - \gl)^{-1} \ge (A_0 - \gl)^{-1}
\eed
for $\gl < \min\{\inf\gs(A_0),\inf\gs(A_\gT)\}$ which yields $A_\gT
\le A_0$.
\end{proof}
\begin{cor}\label{corex}
Let $A$ be a densely defined closed 
symmetric operator in $\gotH$ and let $\{\cH,\Gamma_0,\Gamma_1\}$ be a boundary
triplet for $A^*$ with $\gamma$-field $\gamma(\cdot)$ and Weyl
function $M(\cdot)$. 
Assume that $A$ has finite defect 
and that $A_0=A^*\upharpoonright\ker(\Gamma_0)$ is the Friedrichs
extension. 
Then every selfadjoint extension $A_\Theta$ 
of $A$ in $\gotH$ is semibounded from below and 
\begin{equation*}
\ran\bigl(\gga(\gl)\bigl(\gT - M(\gl)\bigr)^{-1}\bigr) \subseteq \dom(\sqrt{A_\gT - \gl})
\end{equation*}
is  satisfied for all $\gl < \min\{\inf\gs(A_0),\inf\gs(A_\gT)\}$.
\end{cor}

In the next proposition we obtain a
representation of the function $\lambda\mapsto (\Theta-M(\lambda))^{-1}$ in terms
of eigenvalues and eigenfunctions of $A_\Theta$. This representation 
will play an important role in Section~\ref{coupsys}.

\bp\label{V.4}
Let $A$ be a densely defined closed symmetric operator in $\gotH$ and let 
$\{\kH,\gG_0,\gG_1\}$ be a boundary triplet for $A^*$ with Weyl function $M(\cdot)$.
Let $A_\Theta$ be a selfadjoint extension of $A$ corresponding to
$\Theta=\Theta^*\in\widetilde\cC(\cH)$ and assume that $A_0=A^*\upharpoonright\ker(\Gamma_0)$ and 
$A_\Theta$
are semibounded from below, $A_\Theta\leq A_0$, and that the spectrum of $A_\Theta$ is discrete.
Then the $[\cH]$-valued function $\lambda\mapsto (\Theta-M(\lambda))^{-1}$ admits the
representation
\begin{equation}\label{5.15}
\bigl(\Theta-M(\lambda)\bigr)^{-1} = \sum_{k=1}^\infty (\gl_k - \gl)^{-1} (\cdot,\gG_0\psi_k )\gG_0\psi_k,
\quad \lambda\in\rho(A_0)\cap\rho(A_\Theta),
\end{equation}
where 
$\{\gl_k\}$, $k=1,2,\dots$, are the eigenvalues of $A_\Theta$ in increasing order and
$\{\psi_k\}$ are the corresponding eigenfunctions. The
convergence in \eqref{5.15} is understood in the strong sense.
\ep
\begin{proof}
Let $\gl_0 < \min\{\inf\gs(A_0),\inf\gs(A_\gT)\}$ and
let $E_m$, $m \in \bN$, be the orthogonal projection in $\gotH$ onto the
subspace spanned by the eigenfunctions $\{\psi_k\}$, $k=1,\dots,m < \infty$, of $A_\Theta$.
Considerations similar as in 
the proof of Proposition~\ref{II.4} show
\begin{equation*}
\begin{split}
&\Gamma_0 E_m\gamma(\lambda_0)\bigl(\Theta-M(\lambda_0)\bigr)^{-1}\\
&\qquad\qquad=\Gamma_0(A_\Theta-\lambda_0)^{-1/2}E_m\sqrt{A_\Theta-\lambda_0}\,
\gamma(\lambda_0)\bigl(\Theta-M(\lambda_0)\bigr)^{-1}\\
&\qquad\qquad=\overline F_\Theta(\lambda_0)E_m\overline F_\Theta(\lambda_0)^*,
\end{split}
\end{equation*}
where $F_\Theta(\lambda_0)$ is defined as in \eqref{fthe} and $\overline F_\Theta(\lambda_0)\in[\gotH,\cH]$ 
denotes the closure. Hence we have
\begin{equation*}
\lim_{m\rightarrow\infty}\Gamma_0 E_m\gamma(\lambda_0)\bigl(\Theta-M(\lambda_0)\bigr)^{-1}
= \overline F_\Theta(\lambda_0)\overline F_\Theta(\lambda_0)^*=\bigl(\Theta-M(\lambda_0)\bigr)^{-1}
\end{equation*}
in the strong topology.
For $\gl \in \rho(A_0) \cap \rho(A_\Theta)$ we conclude from the representations
\begin{equation*}
\begin{split}
\bigl(\Theta-M(\lambda)\bigr)^{-1}&=\Gamma_0\bigl(\Gamma_0(A_\Theta-\overline\lambda)^{-1}\bigr)^*\\
&=\overline{F}_{\gT}(\gl_0) (A_\Theta - \gl_0)(A_\Theta - \gl)^{-1}\overline{F}_{\gT}(\gl_0)^*
\end{split}
\end{equation*}
and
\begin{equation*}
\Gamma_0 E_m\gamma(\lambda)\bigl(\Theta-M(\lambda)\bigr)^{-1}=
\overline{F}_{\gT}(\gl_0) (A_\Theta - \gl_0)(A_\Theta - \gl)^{-1}E_m\overline{F}_{\gT}(\gl_0)^*
\end{equation*}
that
\begin{equation*}
\lim_{m\rightarrow\infty} 
\Gamma_0 E_m\gamma(\lambda)\bigl(\Theta-M(\lambda)\bigr)^{-1}=\bigl(\Theta-M(\lambda)\bigr)^{-1}
\end{equation*}
in the strong sense for all $\lambda\in\rho(A_0)\cap\rho(A_\Theta)$.

Further, since the resolvent of $A_\Theta$ admits the representation
\bed
(A_\Theta - \gl)^{-1} = \sum^\infty_{k=1} (\gl_k - \gl)^{-1}(\cdot,\psi_k)\psi_k, \quad
\gl \in \rho(A_\Theta),
\eed
where the convergence is in the strong sense, we find
\bed
\Gamma_0 (A_\Theta - \gl)^{-1}E_m = \sum^m_{k=1}(\gl_k - \gl)^{-1}(\cdot,\psi_k)\Gamma_0\psi_k.
\eed
For $\lambda\in\rho(A_0)\cap\rho(A_\Theta)$ the adjoint operator is
given by
\begin{equation*}
\begin{split}
E_m\bigl(\Gamma_0 (A_\Theta - \gl)^{-1}\bigr)^*
&=E_m\bigl(\bigl(\Theta-M(\lambda)\bigr)^{-1}\gamma(\overline\lambda)^*\bigr)^*
=E_m\gamma(\overline\lambda)\bigl(\Theta-M(\overline\lambda)\bigr)^{-1}\\
&=\sum^m_{k=1}(\gl_k - \overline\gl)^{-1}
(\cdot,\Gamma_0\psi_k)\psi_k.
\end{split}
\end{equation*}
Here we have again used \eqref{2.8}, 
$A_0=A^*\upharpoonright\ker(\Gamma_0)$ and $\Gamma_0\gamma(\lambda)=I_\cH$.
Replacing $\lambda$ by $\overline\lambda$ and applying $\Gamma_0$ we obtain from the above formula
the representation
\begin{equation*}
\Gamma_0 E_m\gamma(\lambda)\bigl(\Theta-M(\lambda)\bigr)^{-1}=\sum^m_{k=1}(\gl_k - \gl)^{-1}
(\cdot,\Gamma_0\psi_k)\Gamma_0\psi_k
\end{equation*}
for all $\lambda\in\rho(A_0)\cap\rho(A_\Theta)$.
By the above arguments the left hand side converges in the strong sense to $(\Theta-M(\lambda))^{-1}$. 
Therefore we obtain \eqref{5.15}.
\end{proof}

The special case $\Theta=0\in[\cH]$ will be of particular interest in our further
investigations. In this situation Proposition~\ref{V.4} reads as follows.

\begin{cor}\label{V.14}
Let $A$ be a densely defined closed symmetric operator in $\gotH$ and let 
$\{\kH,\gG_0,\gG_1\}$ be a boundary triplet for $A^*$ with Weyl function $M(\cdot)$.
Assume that $A_0=A^*\upharpoonright\ker(\Gamma_0)$ and 
$A_1=A^*\upharpoonright\ker(\Gamma_1)$ are semibounded from below, 
$A_1\leq A_0$, and that
$\sigma(A_1)$ is discrete.
Then the $[\cH]$-valued function $\lambda\mapsto M(\lambda)^{-1}$ admits the
representation
\begin{equation}\label{5.151}
M(\lambda)^{-1} = \sum_{k=1}^\infty (\gl-\gl_k)^{-1} (\cdot,\gG_0\psi_k)\gG_0\psi_k,
\quad \lambda\in\rho(A_0)\cap\rho(A_1),
\end{equation}
where 
$\{\gl_k\}$, $k=1,2,\dots$, are the eigenvalues of $A_1$ in increasing order,
$\{\psi_k\}$ are the corresponding eigenfunctions, and the
convergence in \eqref{5.151} is understood in the strong sense.
\end{cor}

Proposition~\ref{V.4} and Corollary~\ref{V.14} might suggest that the Weyl function $M$ can be
represented as a convergent series involving the eigenvalues and eigenfunctions of the selfadjoint
operator $A_0$. The following proposition shows that this is 
not possible if $A_0$ is chosen to be the Friedrichs extension.   

\begin{prop}\label{negativ}
Let $A$ be a densely defined closed symmetric operator in $\gotH$ with finite or infinite deficiency indices 
and let $\{\cH,\gG_0,\gG_1\}$ be a boundary triplet for $A^*$.
Assume that $A_0=A^*\upharpoonright\ker(\Gamma_0)$ and 
$A_1=A^*\upharpoonright\ker(\Gamma_1)$ are semibounded, 
that $A_0$ coincides with the Friedrichs extension of $A$ and that 
$\sigma(A_0)$ is discrete.
Then the limit 
\begin{equation*}
\lim_{m\rightarrow\infty}\sum_{k=1}^m (\gl-\mu_k)^{-1} (\cdot,\gG_1\phi_k)\gG_1\phi_k,
\quad \lambda\in\rho(A_0),
\end{equation*}
where 
$\{\mu_k\}$, $k=1,2,\dots$, are the eigenvalues of $A_0$ in increasing order and
$\{\phi_k\}$ are the corresponding eigenfunctions, does not exist. 
\end{prop}

\begin{proof}
We set
\begin{equation}\label{qlambda}
Q(\gl) := \gG_1(A_0 - \gl)^{-1}, \quad \gl \in \rho(A_0),
\end{equation}
and
\begin{equation*}
G(\gl) := \gG_1Q(\overline{\gl})^* = 
\Gamma_1\bigl(\gG_1(A_0 - \overline\gl)^{-1}\bigr)^*, \quad \gl \in \rho(A_0).
\end{equation*}
Taking into account the relation
\begin{equation*}
(A_1 - \gl)^{-1} = (A_0 - \gl)^{-1} -
\gga(\gl)M(\gl)^{-1}\gga(\overline{\gl})^*,
\qquad \gl \in \rho(A_0) \cap \rho(A_1),
\end{equation*}
and \eqref{2.3A} we find
\begin{equation*}
Q(\gl) = \gga(\overline{\gl})^*\quad\text{and}\quad G(\gl) = M(\gl)\qquad \gl \in \rho(A_0)
\cap \rho(A_1).
\end{equation*}
Let $m\in\dN$, let $E_m$ be the projection onto the subspace 
spanned by the eigenfunctions $\{\phi_k\}$, $k=1,\dots,m$, 
and define 
\begin{equation*}
Q^m(\gl) := Q(\gl)E_m\quad\text{and}\quad G^m(\gl) := \gG_1E_m Q(\overline{\gl})^*,
 \quad \gl \in \rho(A_0).
\end{equation*}
With the help of 
\begin{equation*}
(A_0-\overline\lambda)^{-1}=\sum_{k=1}^\infty(\mu_k-\overline\lambda)^{-1}(\cdot,\phi_k)\phi_k
\end{equation*}
and \eqref{qlambda} we find the representation
\begin{equation*}
G^m(\gl) = \sum^m_{k=1} (\mu_k-\lambda )^{-1}\left(\cdot,\Gamma_1\phi_k\right)\Gamma_1\phi_k, 
\quad \gl \in \rho(A_0) \cap \rho(A_1),
\end{equation*}
and on the other hand
\begin{equation*}
G^m(\gl) = Q^m(\gl)(A_0 - \gl)E_m Q(\overline{\gl})^*= \gga(\overline{\gl})^*(A_0 - \gl)E_m\gga(\gl)
\end{equation*}
for $\lambda\in\rho(A_0)\cap \rho(A_1)$.

Let $\gl \in \bR$, $\gl < \min\{\inf\gs(A_0),\inf\gs(A_1)\}$, and
assume that there is an element $\eta \in \kH$ such that
the limit
\begin{equation}\la{6.23}
\lim_{m\to\infty} G^m(\gl)\eta =
\lim_{m\to\infty}\sum^m_{k = 1}(\mu_k - \gl)^{-1}\left(\eta,\Gamma_1\phi_k\right)\Gamma_1\phi_k 
\end{equation}
exists. Since for $h := \gga(\gl)\eta\in\cN_\lambda=\ker(A^*-\lambda)$
\begin{equation*}
\left(G^m(\gl)\eta,\eta\right)= \bigl((A_0
  -\gl)E_m\gga(\gl)\eta,\gga(\gl)\eta\bigr) = \left\|\sqrt{A_0 - \gl}E_mh\right\|^2
\end{equation*}
we obtain from \eqref{6.23} that the limit $\lim_{m\to\infty}\Vert \sqrt{A_0 - \gl}E_mh\Vert$ 
exists and is finite. Therefore there is a subsequence $\{m_n\}$, $n\in\dN$, such that
\begin{equation*}
g := \wlim_{n\to\infty}\sqrt{A_0 - \gl}\,E_{m_n}h\quad\text{and}\quad \lim_{n\to\infty}E_{m_n}h = h.
\end{equation*}
Hence we conclude $h \in \dom(\sqrt{A_0 - \gl})$ and $g = \sqrt{A_0 -\gl}\,h$.
But according to \cite[Lemma 2.1]{AS1} we have $\dom(\sqrt{A_0 - \gl}) \cap \kN_\gl = \{0\}$, so that $h = 0$
and therefore $\eta = 0$. 
\end{proof}

\section{Scattering theory and representation of $S$ and $R$-matrices}\label{scatsec}

Let $A$ be a densely defined closed simple symmetric operator in the separable Hilbert space
$\gotH$
and assume that the deficiency indices of $A$ coincide and are finite, $n_+(A)=n_-(A)<\infty$.
Let $\{\kH,\gG_0,\gG_1\}$
be a boundary triplet for $A^*$, $A_0=A^*\upharpoonright\ker(\Gamma_0)$, 
and let $A_\Theta$ be a selfadjoint extension of $A$
which corresponds to a selfadjoint relation $\Theta\in\widetilde\cC(\cH)$. Note that $\dim\cH=n_\pm(A)$ is
finite. Let $P_\op$
be the orthogonal projection in $\kH$ onto the subspace $\cH_\op:=\dom(\Theta)$ and
decompose $\Theta$ as in \eqref{thetadeco}, $\Theta=\Theta_\op\oplus\Theta_\infty$ with respect
to $\cH_\op\oplus\cH_\infty$. 
The Weyl function $M(\cdot)$
corresponding to $\{\kH,\gG_0,\gG_1\}$ is
a matrix-valued Nevanlinna function and the same holds for 
\begin{equation}\label{nthetaop}
N_\Theta(\lambda):=\bigl(\Theta-M(\lambda)\bigr)^{-1}=\bigl(\Theta_\op-M_\op(\lambda)\bigr)^{-1}P_\op,
\quad\lambda\in\dC\backslash\dR,
\end{equation}
where $M_\op(\lambda)=P_\op M(\lambda)P_\op$, cf. \cite[page 137]{LT77}.
We will in general not distinguish between the orthogonal projection onto $\cH_\op$ and  the
canonical embedding of $\cH_\op$ into $\cH$.
By Fatous theorem (see \cite{Don,Gar}) the limits
\begin{equation*}
M(\gl + i0) := \lim_{\epsilon\to+0}M(\gl + i\epsilon)
\end{equation*}
and
\begin{equation*}
N_\Theta(\lambda+i0):=\lim_{\epsilon\to+0}\bigl(\Theta-M(\lambda+i\epsilon)\bigr)^{-1}
\end{equation*}
from the upper half-plane exist for a.e. $\gl \in \bR$. We denote the set of real points
where the limits exist by $\gS^M$ and $\gS^{N_\Theta}$, respectively, and we agree to use a similar
notation for arbitrary scalar and matrix-valued Nevanlinna functions. It is not difficult to see that
\begin{equation*}
N_{\Theta}(\lambda+i0)=\bigl(\Theta-M(\lambda+i0)\bigr)^{-1}
=\bigl(\Theta_\op-M_\op(\lambda+i0)\bigr)^{-1}P_\op,
\end{equation*}
holds for all $\lambda\in\Sigma^M\cap\Sigma^{N_\Theta}$ and 
that $\dR\backslash(\Sigma^M\cap\Sigma^{N_\Theta})$ 
has Lebesgue measure zero, cf. \cite[$\S 2.3$]{BMN2}.

Since $\dim\cH$ is finite by \eqref{2.8}
\begin{equation*}
\dim\Bigl(\ran\bigl((A_\Theta-\lambda)^{-1}-(A_0-\lambda)^{-1}\bigr)\Bigr)
<\infty,\quad\lambda\in\rho(A_\Theta)\cap\rho(A_0),
\end{equation*}
and therefore the pair $\{A_\gT,A_0\}$ performs a so-called
{\it complete scattering system}, that is, the {\it wave operators}
\begin{equation*}
W_\pm(A_\Theta,A_0) := \slim_{t\to\pm\infty}e^{itA_\Theta}e^{-itA_0}P^{ac}(A_0),
\end{equation*}
exist and their ranges coincide with the absolutely continuous
subspace $\gotH^{ac}(A_\Theta)$ of $A_\Theta$, cf. \cite{BW,Ka1,Wei1,Y}.
 $P^{ac}(A_0)$ denotes the orthogonal
projection onto the absolutely continuous subspace $\gotH^{ac}(A_0)$
of $A_0$.
The {\it scattering operator} $S_\Theta$ of the {\it scattering system}
$\{A_\Theta,A_0\}$ is then defined by
\begin{equation*}
S_\Theta:= W_+(A_\Theta,A_0)^*W_-(A_\Theta,A_0).
\end{equation*}
If we regard the scattering operator as an operator in $\gotH^{ac}(A_0)$,
then $S_\Theta$ is unitary, commutes with the absolutely continuous part
\begin{equation*}
A^{ac}_0:=A_0\upharpoonright \dom(A_0)\cap\gotH^{ac}(A_0)
\end{equation*}
of $A_0$ and it follows
that $S_\Theta$ is unitarily equivalent to a multiplication operator
induced by a family $\{S_\Theta(\lambda)\}$ of unitary operators in
a spectral representation of $A_0^{ac}$, see e.g. \cite[Proposition 9.57]{BW}.
This family is called
the {\it scattering matrix} of the scattering system $\{A_\Theta,A_0\}$. 

In \cite{BMN1} a representation theorem for the scattering matrix 
$\{S_\Theta(\lambda)\}$ in terms of the Weyl
function $M(\cdot)$ was proved, which is of similar type as 
Theorem~\ref{scattering} below. We will make use of the notation
\begin{equation}\label{hm}
\kH_{M(\gl)} := \ran\bigl(\imag(M(\gl))\bigr),\qquad\lambda\in\Sigma^M,
\end{equation}
and we will usually regard $\cH_{M(\lambda)}$ as a subspace of $\cH$.
The orthogonal projection onto $\cH_{M(\lambda)}$ will be denoted by $P_{M(\lambda)}$.
Note that for $\lambda\in\rho(A_0)\cap\dR$ the Hilbert space $\kH_{M(\gl)}$
is trivial by \eqref{mlambda}. The family $\{P_{M(\lambda)}\}_{\lambda\in\Sigma^M}$ of orthogonal
projections in $\cH$ onto $\cH_{M(\lambda)}$, $\lambda\in\Sigma^M$, is measurable and
defines an orthogonal projection in the Hilbert space $L^2(\dR,d\lambda,\cH)$. The range of
this projection is denoted by $L^2(\dR,d\lambda,\cH_{M(\lambda)})$. 
Let $P_\op$ and $M_\op(\lambda)=P_\op M(\lambda)P_\op$, $\lambda\in\Sigma^M$, be as above.
For each $\lambda\in\Sigma^M$ the space $\kH_{M(\gl)}$ will also be written as the orthogonal sum of 
\begin{equation*}
\cH_{M_\op(\lambda)}=\ran\bigl(\imag(M_\op(\gl))\bigr)
\end{equation*}
and
\begin{equation*}
\cH_{M_\op(\lambda)}^\bot:=\cH_{M(\lambda)}\ominus\cH_{M_\op(\lambda)}= \ker\bigl(\imag(M_\op(\gl))\bigr).
\end{equation*}

The following theorem is a variant of \cite[Theorem 3.8]{BMN1}. The essential advantage here
is, that the particular form of the scattering matrix $\{S_\Theta(\lambda)\}$ immediately shows that the 
multivalued part of the selfadjoint parameter $\Theta$ has no influence on the scattering matrix.

\begin{thm}\label{scattering}
Let $A$ be a densely defined closed simple symmetric operator with equal
finite deficiency indices in the separable Hilbert space $\gotH$ and
let $\Pi= \{\kH,\Gamma_0,\Gamma_1\}$ be a boundary triplet for $A^*$
with corresponding Weyl function $M(\cdot)$. Furthermore, let
$A_0=A^*\!\upharpoonright\ker(\Gamma_0)$  and
let $A_\Theta$ be a selfadjoint extension of $A$ which corresponds to 
$\Theta=\Theta_\op\oplus\Theta_\infty\in\widetilde\cC(\cH)$
via \eqref{bij}.
Then the following holds.
\begin{itemize}
\item [{\rm (i)}] The absolutely continuous part
$A^{ac}_0$ of $A_0$ is unitarily equivalent to the multiplication operator
with the free variable in $L^2(\bR,d\gl,\kH_{M(\gl)})$.
\item [{\rm (ii)}] With respect to the 
decomposition $\cH_{M(\lambda)}=\cH_{M_\op(\lambda)}\oplus\cH_{M_\op(\lambda)}^\bot$
the scattering matrix $\{S_\Theta(\gl)\}$
of the complete scattering system $\{A_\Theta,A_0\}$ in $L^2(\bR,d\gl,\kH_{M(\gl)})$ is given by
\begin{displaymath}
S_\Theta(\lambda)=\begin{pmatrix} 
S_{\Theta_\op}(\lambda) & 0 \\ 0 & I_{\cH_{M_\op(\lambda)}^\bot} 
\end{pmatrix}
\in\bigl[ \cH_{M_\op(\lambda)}\oplus\cH_{M_\op(\lambda)}^\bot \bigr],
\end{displaymath}
where
\begin{displaymath}
S_{\Theta_\op}(\lambda)=I_{\cH_{M_\op(\lambda)}} +
2i\sqrt{\imag(M_\op(\gl))}\bigl(\Theta_\op-M_\op(\gl)\bigr)^{-1}
\sqrt{\imag(M_\op(\gl))}
\end{displaymath}
and $\gl \in\Sigma^M\cap\Sigma^{N_\Theta}$, $M_\op(\gl):=M_\op(\gl +
i0)$.
\end{itemize}
\end{thm}

\begin{proof}
Assertion (i) was proved in \cite[Theorem 3.8]{BMN1} and moreover it was shown that
the scattering matrix $\{\widetilde S_\Theta(\gl)\}$
of the complete scattering system $\{A_\Theta,A_0\}$ in $L^2(\bR,d\gl,\kH_{M(\gl)})$ has the form
\begin{equation*}
\widetilde S_\Theta(\gl)=
I_{\cH_{M(\lambda)}} +
2i\sqrt{\imag(M(\gl))}\bigl(\Theta-M(\gl)\bigr)^{-1}
\sqrt{\imag(M(\gl))}\in [\cH_{M(\lambda)}]
\end{equation*}
for all $\gl \in\Sigma^M\cap\Sigma^{N_\Theta}$. With the help of \eqref{nthetaop} this becomes
\begin{equation*}
\widetilde S_\Theta(\gl)=I_{\cH_{M(\lambda)}} +
2i\sqrt{\imag(M(\gl))}P_\op\bigl(\Theta_\op-M_\op(\gl)\bigr)^{-1}P_\op
\sqrt{\imag(M(\gl))}.
\end{equation*}
From the polar decomposition of $\sqrt{\imag(M(\gl))}P_\op$, $\lambda\in\Sigma^M$, we obtain a family
of isometric mappings $V(\lambda)$, $\lambda\in\Sigma^M$, 
from $\cH_{M_\op(\lambda)}$ onto $\ran(\sqrt{\imag (M(\gl))}P_\op)$ defined by
\begin{equation*}
V(\lambda)\sqrt{\imag(M_\op(\lambda))}\,x:=\sqrt{\imag (M(\gl))}\,P_\op x
\end{equation*}
and we extend $V(\lambda)$ to a family $\widetilde V(\lambda)$ 
of unitary mappings in $\cH_{M(\lambda)}$. Note that $\widetilde V(\lambda)$
maps $\ker(\sqrt{\imag(M_\op(\lambda))})$ isometrically onto $\ker(P_\op\sqrt{\imag(M(\gl))})$.
It is not difficult to see that the scattering matrix
\begin{equation*}
S_\Theta(\lambda):=
\widetilde V(\lambda)^*\widetilde S_\Theta(\lambda)\widetilde
V(\lambda),
\qquad\lambda\in\Sigma^M\cap\Sigma^{N_\Theta},
\end{equation*}
with respect to the decomposition $\cH_{M(\lambda)}=\cH_{M_\op(\lambda)}\oplus\cH_{M_\op(\lambda)}^\bot$
is of the form as in assertion (ii).
\end{proof}
 
We point out that the scattering matrix $\{S_\Theta(\gl)\}$
of the complete scattering system $\{A_\Theta,A_0\}$ is defined for a.e. $\lambda\in\dR$ and that 
in Theorem~\ref{scattering}(ii) a special representative of the corresponding equivalence class was chosen.
We also note that the operator $\sqrt{\imag(M_\op(\gl))}$ is regarded as an operator in $\kH_{M_\op(\gl)}$.

Next we introduce the {\it $R$-matrix} $\{R_\Theta(\lambda)\}$ of 
the scattering system $\{A_\Theta,A_0\}$ in accordance with 
Blatt and Weiskopf \cite{BW1},
\begin{equation}\label{rmatrix}
R_\Theta(\lambda):=
i\bigl(I_{\kH_{M(\gl)}}-S_\Theta(\lambda)\bigr)\bigl(I_{\kH_{M(\gl)}}+S_\Theta(\lambda)\bigr)^{-1}
\end{equation}
for all $\lambda\in\Sigma^M\cap\Sigma^{N_\Theta}$ 
satisfying $-1\in\rho(S_\Theta(\lambda))$. Since $S_\Theta(\lambda)$ is unitary
it follows that $R_\Theta(\lambda)$ is a selfadjoint matrix.
Note also that
\begin{equation}\label{srrel}
S_\Theta(\lambda)=\bigl(iI_{\kH_{M(\gl)}}  -R_\Theta(\lambda)\bigr)
\bigl(i I_{\kH_{M(\gl)}}+ R_\Theta(\lambda) \bigr)^{-1}
\end{equation}
holds for all real $\lambda$ where $R_\Theta(\lambda)$ is defined.

The next theorem is of similar flavor as Theorem~\ref{scattering}. 
We express the $R$-matrix of the scattering system 
$\{A_\Theta,A_0\}$ in terms of the Weyl function $M(\cdot)$ and the
selfadjoint parameter $\Theta\in\widetilde\cC(\cH)$. 
Again we make use of the special space decomposition which shows that 
the ``pure'' relation part $\Theta_\infty$ has
no influence on the $R$-matrix.

\begin{thm}\label{rzmat}
Let $A$ be a densely defined closed simple symmetric operator with
equal finite deficiency indices in the separable Hilbert space $\gotH$ and
let $\Pi= \{\kH,\Gamma_0,\Gamma_1\}$ be a boundary triplet for $A^*$
with corresponding Weyl function $M(\cdot)$. Furthermore, let
$A_0=A^*\!\upharpoonright\ker(\Gamma_0)$  and
let $A_\Theta$ be a selfadjoint extension of $A$ corresponding to $\Theta\in\widetilde\cC(\cH)$.
Then for all $\lambda\in\Sigma^M\cap\Sigma^{N_\Theta}$ with 
\begin{equation*}
\ker\bigl(\Theta_\op-\real(M_\op(\gl))\bigr)=\{0\}
\end{equation*}
the $R$-matrix of $\{A_\Theta,A_0\}$ is given by 
\begin{displaymath}
R_\Theta(\gl) = \begin{pmatrix} \sqrt{\imag(M_\op(\gl))}\bigl(\Theta_\op-\real (M_\op(\gl))\bigr)^{-1}
\sqrt{\imag(M_\op(\gl))} & 0\\ 0 & 0\end{pmatrix},
\end{displaymath}
with respect to $\cH_{M(\lambda)}=\cH_{M_\op(\lambda)}\oplus\cH_{M_\op(\lambda)}^\bot$, 
where $M_\op(\lambda)=M_\op(\lambda+i0)$.
\end{thm}

\begin{proof}
It follows immediately from the definition \eqref{rmatrix} and the representation of
the scattering matrix in Theorem~\ref{scattering} (ii), 
that the $R$-matrix of $\{A_\Theta,A_0\}$ is a diagonal block
matrix with respect to the space decomposition 
$\cH_{M(\lambda)}=\cH_{M_\op(\lambda)}\oplus\cH_{M_\op(\lambda)}^\bot$ 
and that the restriction of $R_\Theta(\lambda)$ to
$\cH_{M_\op(\lambda)}^\bot$ 
is identically equal to zero.

Moreover, for every $\lambda\in\Sigma^M\cap\Sigma^{N_\Theta}$ it 
follows from the representation of the scattering matrix that
\begin{equation*}
\begin{split}
&\sqrt{\imag(M_\op(\lambda))}\bigl(I_{\cH_{M_\op(\lambda)}}+S_{\Theta_\op}(\lambda)\bigr)\\
&\quad=2\bigl\{I_{\cH_{M_\op(\lambda)}}+i\imag(M_\op(\lambda))\bigl( \Theta_\op-M_\op(\gl)\bigr)^{-1}\bigr\}
\sqrt{\imag(M_\op(\gl))}\\
&\quad=2\bigl(\Theta_\op-\real(M_\op(\lambda))\bigr)\bigl( \Theta_\op-M_\op(\gl)\bigr)^{-1}\sqrt{\imag(M_\op(\gl))}
\end{split}
\end{equation*}
holds.
If $\lambda\in\Sigma^M\cap\Sigma^{N_\Theta}$ is such 
that $\Theta_\op-\real(M_\op(\lambda))$ is invertible, then we obtain
\begin{equation*}
\begin{split}
&\sqrt{\imag(M_\op(\lambda))}\bigl(I_{\cH_{M_\op(\lambda)}}+S_{\Theta_\op}(\lambda)\bigr)^{-1}\\
&\quad\quad\quad=\frac{1}{2}\bigl( \Theta_\op-M_\op(\gl)\bigr)
\bigl(\Theta_\op-\real(M_\op(\lambda))\bigr)^{-1}
\sqrt{\imag(M_\op(\gl))},
\end{split}
\end{equation*}
so that

\begin{equation*}
\begin{split}
2i\bigl( \Theta_\op-M_\op(\gl)\bigr)^{-1}& \sqrt{\imag(M_\op(\lambda))}\bigl(I_{\cH_{M_\op(\lambda)}}
+S_{\Theta_\op}(\lambda) \bigr)^{-1}\\
&=i\bigl(\Theta_\op-\real(M_\op(\lambda))\bigr)^{-1}\sqrt{\imag(M_\op(\gl))}.
\end{split}
\end{equation*}
Finally multiplication by $-\sqrt{\imag(M_\op(\lambda))}$ from the
left gives 
\begin{equation*}
\begin{split}
\bigl(I_{\cH_{M_\op(\lambda)}}-& S_{\Theta_\op}(\lambda)\bigr) 
\bigl(I_{\cH_{M_\op(\lambda)}}+ S_{\Theta_\op}(\lambda)  \bigr)^{-1}\\
&=-i\sqrt{\imag(M_\op(\lambda))}\bigl(\Theta_\op-\real(M_\op(\lambda))\bigr)^{-1}\sqrt{\imag(M_\op(\gl))}
\end{split}
\end{equation*}
so that assertion (i) follows immediately from the definition of the $R$-matrix in \eqref{rmatrix}.
\end{proof}

\section{Scattering in coupled systems}\label{coupsys}

Let $\sH$ and $\sK$ be separable Hilbert spaces and let
$A$ and $T$ be densely defined closed simple symmetric operators in  
$\gotH$ and $\gotK$, respectively. We assume 
that the deficiency indices of $A$ and $T$ coincide and are finite,
\begin{equation*}
n:=n_+(A)=n_-(A)=n_+(T)=n_-(T)<\infty.
\end{equation*}
Then there exist boundary triplets $\{\cH,\Gamma_0,\Gamma_1\}$ and
$\{\cH,\gY_0,\gY_1\}$  for the adjoint operators
$A^*$ and $T^*$, respectively, with fixed selfadjoint extensions
\begin{equation}\label{a0t0} 
A_0:=A^*\upharpoonright\ker(\Gamma_0)\qquad\text{and}\qquad T_0:=T^*\upharpoonright\ker(\gY_0)
\end{equation}
in $\gotH$ and $\gotK$, respectively, and $\dim\cH=n$. The Weyl functions of $\{\cH,\Gamma_0,\Gamma_1\}$
and $\{\cH,\gY_0,\gY_1\}$ will be denoted by $M(\cdot)$ and $\tau(\cdot)$, respectively.
Besides the spaces $\cH_{M(\lambda)}$, $\lambda\in\Sigma^M$, 
(see \eqref{hm}) we will make use of the finite dimensional spaces 
\begin{equation*}
\cH_{\tau(\lambda)}=\ran\bigl(\imag(\tau(\lambda+i0))\bigr)
,\qquad\lambda\in\Sigma^\tau,
\end{equation*}
and 
\begin{equation*}
\cH_{(M+\tau)(\lambda)}=\ran\bigl(
\imag\bigl((M+\tau)(\lambda+i 0)\bigr)\bigr),
\quad\lambda\in\Sigma^{M+\tau}\supset\bigl(\Sigma^M\cap\Sigma^\tau\bigr).
\end{equation*}

In the following theorem we calculate the $S$ and $R$-matrix of a special scattering system 
$\{\widetilde L,L_0\}$ in $\gotH\oplus\gotK$ 
in terms of the Weyl functions $M$ and $\tau$. Theorem~\ref{srzthm} is in principle a consequence 
of Theorem~\ref{scattering} and Theorem~\ref{rzmat}, cf. \cite[Theorem~4.5]{BMN2}. 
We note that the coupling procedure in the first part of the theorem is similar to the one in \cite{DHMS00}.

\begin{thm}\label{srzthm}
Let $A$, $\{\cH,\Gamma_0,\Gamma_1\}$, $M(\cdot)$ and $T$, $\{\cH,\gY_0,\gY_1\}$, $\tau(\cdot)$ be as above. 
Then the following holds.
\begin{itemize}
\item [{\rm (i)}] The pair $\{\widetilde L,L_0\}$, where $L_0:=A_0\oplus T_0$ and 
\begin{equation}\la{athe}
\widetilde L=A^*\oplus T^*\!\upharpoonright \left\{
f \oplus g \in\dom(A^*\oplus T^*):
\begin{array}{l}
\Gamma_0 f-\gY_0 g = 0\\
\Gamma_1 f+\gY_1 g=0
\end{array}
\right\},
\end{equation}
forms a complete scattering system in the Hilbert space $\gotH\oplus\gotK$ and $L_0^{ac}$ 
is unitarily equivalent to the multiplication operator
with the free variable in $L^2(\bR,d\gl,\kH_{M(\lambda)}\oplus
\kH_{\tau(\lambda)})$.
\item [{\rm (ii)}] 
With respect to the decomposition 
\begin{equation}\label{mschldeco}
\cH_{(M+\tau)(\lambda)}\oplus \cH_{(M+\tau)(\lambda)}^\bot 
\end{equation} 
of $\kH_{M(\lambda)}\oplus \kH_{\tau(\lambda)}$ 
the scattering matrix $\{\widetilde S(\gl)\}$ of 
$\{\widetilde L,L_0\}$ is given by
\begin{displaymath}
\widetilde S(\gl) = \begin{pmatrix} S(\lambda)  & 0 \\ 0 & I_{\cH_{(M+\tau)(\lambda)}^\bot} \end{pmatrix}\in
\bigl[ \cH_{(M+\tau)(\lambda)}\oplus \cH_{(M+\tau)(\lambda)}^\bot  \bigr],
\end{displaymath}
where
\begin{equation*}
\begin{split}
&S(\lambda)=I_{\cH_{(M+\tau)(\lambda)}} \\
&\quad -
2i\sqrt{\imag(M(\lambda)+\tau(\gl))}\bigl(M(\gl)+\tau(\gl)\bigr)^{-1}
\sqrt{\imag(M(\lambda)+\tau(\gl))}
\end{split}
\end{equation*}
and $\gl \in\Sigma^M\cap\Sigma^{\tau}\cap\Sigma^{(M+\tau)^{-1}}$, $M(\gl):=M(\gl + i0)$, 
$\tau(\lambda)=\tau(\lambda+i0)$.
\item [{\rm (iii)}] For all  $\gl \in\Sigma^M\cap\Sigma^{\tau}\cap\Sigma^{(M+\tau)^{-1}}$ with 
$\ker(\real(M(\gl)+\tau(\gl)))=\{0\}$
the $R$-matrix of 
$\{\widetilde L,L_0\}$ is given by
{\small
\bead
\lefteqn{
R(\gl) = }\\
& &\begin{pmatrix} -\sqrt{\imag(M(\gl)+\tau(\gl))}\bigl(\real(M(\gl)+\tau(\gl))\bigr)^{-1}
\sqrt{\imag(M(\gl)+\tau(\gl))} & 0\\ 0 & 0\end{pmatrix}
\eead
}
with respect to the decomposition \eqref{mschldeco}.
\end{itemize}
\end{thm}

\begin{proof}
(i) Let $L:=A\oplus T$, so that $L$ is a densely defined closed simple symmetric operator 
in the Hilbert space $\gotH\oplus\gotK$. Clearly, $L$ has deficiency indices $n_\pm(L)=2n$, and 
it is easy to see that $\{\widetilde\cH,\widetilde\Gamma_0,\widetilde\Gamma_1\}$, where
\begin{equation*}
\widetilde\Gamma_0 (f\oplus g):=\begin{pmatrix}\Gamma_0f\\\gY_0g\end{pmatrix},\quad
\widetilde\Gamma_1 (f\oplus g):=\begin{pmatrix}\Gamma_1f\\\gY_1g\end{pmatrix}
\quad\text{and}\quad \widetilde\cH:=\cH\oplus\cH, 
\end{equation*}
$f\in\dom(A^*)$, $g\in\dom(T^*)$, 
is a boundary triplet for the adjoint operator $L^*=A^*\oplus T^*$ in $\gotH\oplus\gotK$.
Together with the selfadjoint operators 
$A_0$ and $T_0$ from \eqref{a0t0} 
we obviously have
\begin{equation*}
L_0:=L^*\upharpoonright\ker(\widetilde\Gamma_0)=A_0\oplus T_0.
\end{equation*}
It is not difficult to verify that
\begin{equation}\label{wtt}
\widetilde\Theta:=\left\{\begin{pmatrix} (x,x)^\top\\ (y,-y)^\top\end{pmatrix}: x,y\in\cH\right\}\in
\widetilde\cC\bigl(\cH\oplus\cH\bigr)
\end{equation}
is a selfadjoint relation in $\widetilde\cH$ and that the corresponding selfadjoint 
extension $L^*\upharpoonright\widetilde\Gamma^{(-1)}\widetilde\Theta$ in $\gotH\oplus\gotK$ via \eqref{bij} 
coincides with the operator $\widetilde L$ in \eqref{athe}, cf. \cite{BMN2}.
Since $L$ has finite deficiency indices, $\widetilde L$ is finite rank perturbation of $L_0$
in resolvent sense (cf. Theorem~\ref{resthm} and Section~\ref{scatsec}), 
and hence $\{\widetilde L,L_0\}$ is a complete scattering system
in $\gotH\oplus\gotK$. Moreover, as the Weyl function $\widetilde M(\cdot)$ 
of $\{\widetilde\cH,\widetilde\Gamma_0,\widetilde\Gamma_1\}$
is given by 
\begin{equation}\label{mtasd}
\widetilde M(\lambda)=\begin{pmatrix} M(\lambda) & 0\\ 0 & \tau (\lambda) \end{pmatrix},
\qquad\lambda\in\rho(L_0),
\end{equation}
it follows from Theorem~\ref{scattering} (i) that the absolutely continuous part $L_0^{ac}$ of $L_0$
is unitarily equivalent to the multiplication operator with the free variable 
in $L^2(\bR,d\gl,\kH_{\widetilde M(\lambda)})=L^2(\bR,d\gl,\kH_{M(\lambda)}\oplus \kH_{\tau(\lambda)})$.

\vskip 0.2cm\noindent
(ii)-(iii)
Note that the operator part $\widetilde\Theta_\op$ of the selfadjoint relation $\widetilde\Theta$
in \eqref{wtt} is defined on
\begin{equation*}
\widetilde\cH_\op:=\dom(\widetilde\Theta)=\bigr\{(x,x)^\top:x\in\cH\bigl\}
\end{equation*} 
and that $\widetilde\Theta_\op=0\in [\widetilde\cH_\op]$,
cf. \eqref{thetadeco}. 
Next we will calculate the $[\widetilde\cH_\op]$-valued function 
$\widetilde M_\op(\cdot)$, 
and in order to avoid possible confusion we will distinguish
between embeddings and projections here. The canonical embedding of 
$\widetilde\cH_\op$ into $\cH\oplus\cH$ is given by 
\begin{equation*}
\iota_\op:\widetilde\cH_\op\rightarrow\cH\oplus\cH,\qquad 
y\mapsto\frac{1}{\sqrt{2}}\begin{pmatrix}y\\ y\end{pmatrix},
\end{equation*}
and the adjoint $\iota^*_\op \in[\cH\oplus\cH,\widetilde\cH_\op]$ is
the orthogonal projection $P_\op$ 
from $\cH\oplus\cH$ onto $\widetilde\cH_\op$, $P_\op(u\oplus v)=\tfrac{1}{\sqrt{2}}(u+v)$. Then we obtain 
\begin{equation*}
\widetilde M_\op(\lambda)=P_\op\widetilde M(\lambda)\iota_\op=\frac{1}{2}\bigl(M(\lambda)+\tau(\lambda)\bigr),
\qquad\lambda\in\rho(L_0),
\end{equation*}
from \eqref{mtasd}. Now the assertions (ii) and (iii)
follow easily from Theorem~\ref{scattering} (ii) and 
Theorem~\ref{rzmat}, respectively.
\end{proof}

The case that the operator $A_0$ has discrete spectrum is of particular importance in several
applications. In this situation Theorem~\ref{srzthm} reduces to the following corollary.

\begin{cor}\label{srzthmcor}
Let the assumptions and $\{\widetilde L,L_0\}$ 
be as in Theorem~\ref{srzthm} and assume, in addition, that $\sigma(A_0)$ is discrete.
Then the following holds.
\begin{itemize}
\item [{\rm (i)}] $L_0^{ac}$ 
is unitarily equivalent to the multiplication operator
with the free variable in $L^2(\bR,d\gl,\kH_{\tau(\lambda)})$
\item [{\rm (ii)}] 
The scattering matrix $\{S(\gl)\}$ of 
$\{\widetilde L,L_0\}$ in $L^2(\bR,d\gl,\kH_{\tau(\lambda)})$ is given
by
\begin{displaymath}
S(\gl) = I_{\kH_{\tau(\lambda)}} -
2i\sqrt{\imag(\tau(\gl))}\bigl(M(\gl)+\tau(\gl)\bigr)^{-1}
\sqrt{\imag(\tau(\gl))}
\end{displaymath}
for $\gl \in\Sigma^M\cap\Sigma^{\tau}\cap\Sigma^{(M+\tau)^{-1}}$,
where $M(\gl):=M(\gl + i0)$, $\tau(\lambda)=\tau(\lambda+i0)$.
\item [{\rm (iii)}] For all  $\gl \in\Sigma^M\cap\Sigma^{\tau}\cap\Sigma^{(M+\tau)^{-1}}$ with 
$\ker(M(\gl)+\real(\tau(\gl)))=\{0\}$
the $R$-matrix of 
$\{\widetilde L,L_0\}$ is given by
\begin{displaymath}
R(\gl) = -\sqrt{\imag(\tau(\gl))}\bigl(M(\gl)+\real(\tau(\gl))\bigr)^{-1}
\sqrt{\imag(\tau(\gl))} 
\end{displaymath}
\end{itemize}
\end{cor}

\begin{proof}
The assumption $\sigma(A_0)=\sigma_p(A_0)$ implies $\imag(M(\lambda))=\{0\}$ 
for all $\lambda\in\Sigma^M$. Therefore 
\begin{equation*}
\cH_{(M+\tau)(\lambda)}=\cH_{\tau(\lambda)}\quad\text{and}\quad
\cH_{M(\lambda)}=\{0\},\qquad\lambda\in\Sigma^M,
\end{equation*}
and the statements follow immediately from Theorem~\ref{srzthm}.
\end{proof}

From relation \eqref{srrel} we obtain the next corollary. We note that this statement
can be formulated also for the case when $\sigma(A_0)$ is not discrete. However in our
applications we will only make use of the more special variant below.

\begin{cor}\label{cor1}
Let the assumptions be as in Corollary~\ref{srzthmcor}. 
Then for all $\lambda\in \Sigma^M\cap\Sigma^{\tau}\cap\Sigma^{(M+\tau)^{-1}}$
with $\ker(M(\lambda)+\real(\tau(\gl)))=\{0\}$ the scattering matrix $\{S(\gl)\}$ of 
$\{\widetilde L,L_0\}$ admits the representation
\begin{equation*}
\begin{split}
S(\lambda)=\Bigl(iI_{\cH_{\tau(\lambda)}}&+ \sqrt{\imag(\tau(\gl))} 
\bigl(M(\gl)+\real(\tau(\lambda))\bigr)^{-1}
\sqrt{\imag(\tau(\gl))} \Bigr) \\
&\Bigl(iI_{\cH_{\tau(\lambda)}}- \sqrt{\imag(\tau(\gl))} \bigl(M(\gl)+\real(\tau(\lambda))\bigr)^{-1}    
\sqrt{\imag(\tau(\gl))}  \Bigr)^{-1}
\end{split}
\end{equation*}
and, if, in particular, $\real(\tau(\lambda))=0$, then 
\begin{equation*}
\begin{split}
S(\lambda)=\Bigl(iI_{\cH_{\tau(\lambda)}}+& \sqrt{\imag(\tau(\gl))} M(\gl)^{-1}
\sqrt{\imag(\tau(\gl))} \Bigr) \\
&\Bigl(iI_{\cH_{\tau(\lambda)}}- \sqrt{\imag(\tau(\gl))}M(\gl)^{-1}
\sqrt{\imag(\tau(\gl))}  \Bigr)^{-1}.
\end{split}
\end{equation*}
\end{cor}

Our next objective is to express the scattering matrix of the
scattering system 
$\{\widetilde L,L_0\}$ in terms
of the eigenfunctions of a family of selfadjoint extensions of $A$. 
For this let again $\tau(\cdot)$ be the Weyl function
of $\{\cH,\gY_0,\gY_1\}$, let $\mu\in\Sigma^\tau$, and 
let $\{\cH,\Gamma_0,\Gamma_1\}$ be a boundary triplet for
$A^*$ as in the beginning of this section. Then $\real(\tau(\mu))$ is a selfadjoint matrix
in $\cH$ and therefore the operator
\begin{equation}\label{arealt}
A_{-\real(\tau(\mu))}=A^*\upharpoonright\ker\bigl(\Gamma_1+\real(\tau(\mu))\Gamma_0\bigr) 
\end{equation}
is a selfadjoint extension of $A$ in $\gotH$, cf. Proposition~\ref{propo}. Note that by Theorem~\ref{resthm}
a point $\lambda\in\rho(A_0)$
belongs to $\rho(A_{-\real(\tau(\mu))})$ if and only if $0\in\rho(M(\lambda)+\real\tau(\mu))$ holds.
The following corollary is
a reformulation of Proposition~\ref{V.4} in our particular situation.

\begin{cor}\label{asdf}
Let $A$, $\{\cH,\Gamma_0,\Gamma_1\}$, $M(\cdot)$ and $T$, 
$\{\cH,\gY_0,\gY_1\}$, $\tau(\cdot)$ be as above and 
assume $\sigma(A_0)=\sigma_p(A_0)$ and that $A$ is semibounded from below. For each $\mu\in\Sigma^\tau$ with 
$A_{-\real(\tau(\mu))}\leq A_0$ the function $\lambda\mapsto -(\real(\tau(\mu))+M(\lambda))^{-1}$ 
admits the representation
\begin{equation*}
-\bigl(M(\lambda)+\real(\tau(\mu))\bigr)^{-1}=\sum_{k=1}^\infty(\lambda_k[\mu]-\lambda)^{-1}
\bigl(\cdot,\Gamma_0\psi_k[\mu]\bigr)\Gamma_0\psi_k[\mu],
\end{equation*}  
where $\{\lambda_k[\mu]\}$, $k=1,2,\dots$, are the eigenvalues of 
the selfadjoint extension $A_{-\real(\tau(\mu))}$ in increasing
order and $\psi_k[\mu]$ are the corresponding eigenfunctions.
\end{cor}

Setting $\mu=\lambda$ in Corollary~\ref{asdf} and taking into account 
Corollary~\ref{srzthmcor} and Corollary~\ref{cor1} we obtain the following representations 
of the $R$-matrix and scattering matrix  
of $\{\widetilde L,L_0\}$.

\begin{thm}\label{proppo}
Let the assumptions be as in Corollary~\ref{asdf}. Then for all $\lambda\in\Sigma^M\cap
\Sigma^\tau\cap\Sigma^{(M+\tau)^{-1}}$ with $\ker(M(\lambda)+\real(\tau(\gl)))=\{0\}$ and 
$A_{-\real(\tau(\lambda))}\leq A_0$
the $R$-matrix and the scattering matrix of 
$\{\widetilde L,L_0\}$ admit the representations
\begin{equation*}
R(\gl) = \sum_{k=1}^\infty(\lambda_k[\lambda]-\lambda)^{-1}
\bigl(\sqrt{\imag(\tau(\gl))}\cdot,\Gamma_0\psi_k[\lambda]\bigr)\sqrt{\imag(\tau(\gl))}\Gamma_0\psi_k[\lambda]
\end{equation*}
and
\bead
\lefteqn{
S(\lambda)=}\\
& & \hspace{-5mm}
\Bigl(iI_{\cH_{\tau(\lambda)}}- \sum_{k=1}^\infty(\lambda_k[\lambda]-\lambda)^{-1}
\bigl(\sqrt{\imag(\tau(\gl))}\cdot,\Gamma_0\psi_k[\lambda]\bigr)
\sqrt{\imag(\tau(\gl))}\Gamma_0\psi_k[\lambda]\Bigr) \times\\
& & \hspace{-5mm}
\Bigl(iI_{\cH_{\tau(\lambda)}}+ \sum_{k=1}^\infty(\lambda_k[\lambda]-\lambda)^{-1}
\bigl(\sqrt{\imag(\tau(\gl))}\cdot,\Gamma_0\psi_k[\lambda]\bigr)
\sqrt{\imag(\tau(\gl))}\Gamma_0\psi_k[\lambda]\Bigr)^{-1},
\eead
respectively, 
where $\{\lambda_k[\lambda]\}$, $k=1,2,\dots$, are the eigenvalues of the selfadjoint extension 
$A_{-\real(\tau(\lambda))}$ in increasing
order and $\psi_k[\lambda]$ are the corresponding eigenfunctions.
\end{thm}
 
If $\real(\tau(\lambda))=0$ for some $\lambda\in\Sigma^\tau$, 
then the operator $A_{-\real(\tau(\lambda))}$ in \eqref{arealt}
coincides with the selfadjoint operator
$A_1=A^*\upharpoonright\ker(\Gamma_1)$. 
This yields the next corollary.

\begin{cor}\label{0cor}
Let the assumptions be as in Corollary~\ref{asdf}. Then for all $\lambda\in\Sigma^M\cap
\Sigma^\tau\cap\Sigma^{(M+\tau)^{-1}}$ with $\real(\tau(\lambda))=0$, $\ker(M(\lambda))=\{0\}$ 
and 
$A_1\leq A_0$
the $R$-matrix and the scattering matrix of 
$\{\widetilde L,L_0\}$ admit the representations
\begin{equation*}
R(\gl) = \sum_{k=1}^\infty(\lambda_k-\lambda)^{-1}
\bigl(\sqrt{\imag(\tau(\gl))}\cdot,\Gamma_0\psi_k\bigr)\sqrt{\imag(\tau(\gl))}\Gamma_0\psi_k
\end{equation*}
and
\begin{equation*}
\begin{split}
S(\lambda)=&\Bigl(iI_{\cH_{\tau(\lambda)}}- \sum_{k=1}^\infty(\lambda_k-\lambda)^{-1}
\bigl(\sqrt{\imag(\tau(\gl))}\cdot,\Gamma_0\psi_k\bigr)\sqrt{\imag(\tau(\gl))}\Gamma_0\psi_k\Bigr) \\
&\Bigl(iI_{\cH_{\tau(\lambda)}}+ \sum_{k=1}^\infty(\lambda_k-\lambda)^{-1}
\bigl(\sqrt{\imag(\tau(\gl))}\cdot,\Gamma_0\psi_k\bigr)\sqrt{\imag(\tau(\gl))}\Gamma_0\psi_k\Bigr)^{-1},
\end{split}
\end{equation*}
respectively, 
where $\{\lambda_k\}$, $k=1,2,\dots$, are the eigenvalues of the selfadjoint extension 
$A_1$ in increasing
order and $\psi_k$ are the corresponding eigenfunctions.
\end{cor}
\begin{rem}\la{divger}
{\em
The assumption $A_1 \le A_0$ in Corollary \ref{0cor} above is necessary. Indeed, let 
us assume that $A_0 \le A_1$ and that $A_1$ is the Friedrichs
extension. Let us show that in this case the sum
\begin{equation}\label{divergence}
\sum^\infty_{k=1}(\gl_k - \gl)^{-1}(\cdot,\gG_0\psi_k)\gG_0\psi_k
\end{equation}
cannot converge, 
where $\{\gl_k\}$ and $\{\psi_k\}$ are the eigenvalues and
eigenfunctions of $A_1$. For this consider the boundary triplet
$\{\kH,\gG'_0,\gG'_1\}$, $\gG'_0 = \gG_1$ and $\gG'_1 =
-\gG_0$. Obviously $A'_0 = A^*\upharpoonright\ker(\gG'_0) =
A_1$, $A'_1 = A^*\upharpoonright\ker(\gG'_1) = A_0$ and $A'_0$ is the Friedrichs extension. 
By Proposition \ref{negativ} we
obtain that the sum
\bed
\sum^\infty_{k=1}(\gl - \gl_k)^{-1}(\cdot,\gG'_1\psi_k)\gG'_1\psi_k
\eed
diverges, where $\{\gl_k\}$ and $\{\psi_k\}$ are the eigenvalues and
eigenfunctions of $A'_0 = A_1$. Using $\gG'_1 = -\gG_0$ one gets that the sum
\eqref{divergence} diverges.
}
\end{rem}

\section{Scattering systems of differential operators}

In this section we illustrate the general results from 
the previous sections for scattering systems which consist
of regular and singular second order differential operators, see Section~\ref{sturm}. 

\subsection{Coupling of differential operators}\label{coupsturm}

Let the symmetric operators
$A=-\tfrac{1}{2}\tfrac{d}{dx}\tfrac{1}{m}\tfrac{d}{dx}+v$ and 
\begin{equation*}
T=T_l\oplus T_r=\left(-\frac{1}{2}\frac{d}{dx}\frac{1}{m_l}\frac{d}{dx}+v_l\right)\oplus 
\left(-\frac{1}{2}\frac{d}{dx}\frac{1}{m_r}\frac{d}{dx}+v_r\right)
\end{equation*} 
in $\gotH=L^2((x_l,x_r))$ and $\gotK=L^2((-\infty,x_l))\oplus L^2((x_r,\infty))$ and the boundary
triplets $\gP_A=\{\dC^2,\Gamma_0,\Gamma_1\}$ and 
$\gP_T=\{\dC^2,\gY_0,\gY_1\}$ be as in Section~\ref{2.3.1} and
and Section~\ref{2.3.2}, respectively. 
By Theorem~\ref{srzthm}(i) the operator
\begin{equation}\la{buslaev}
\widetilde{L} :=
A^*\oplus T^*\!\upharpoonright
\left\{f \oplus g \in \dom(A^*\oplus T^*):
\ba{l}
\Gamma_0 f- \gY_0 g = 0\\
\Gamma_1 f + \gY_1 g= 0
\ea
\right\}
\end{equation}
is a selfadjoint extension of $L = A\oplus T$ in $\gotH \oplus \gotK$. We can identify
$\gotH \oplus \gotK$ with
\begin{equation*}
L^2((x_l,x_r)) \oplus L^2((-\infty,x_l)) \oplus L^2((x_r,\infty))
\cong L^2(\bR).
\end{equation*} 
The elements $f\oplus g$ in $\gotH \oplus \gotK$, $f\in\gotH$,
$g = g_l \oplus g_r \in \gotK$, will be written as
$f \oplus g_l \oplus g_r$. 
Here the conditions $\Gamma_0 f=\gY_0 g$ and
$\Gamma_1 f=-\gY_1 g$, $f\in\dom(A^*)$, $g\in\dom(T^*)$,
explicitely mean
\bed
g_l(x_l) = f(x_l)
\quad \mbox{and} \quad
f(x_r) = g_r(x_r),
\eed
and
\bed
\left(\frac{1}{m}f'\right)(x_l) = \left(\frac{1}{m_l}g_l'\right)(x_l)  
\quad \mbox{and} \quad
\left(\frac{1}{m}f'\right)(x_r) = \left(\frac{1}{m_r}g_r'\right)(x_r).
\eed
Hence the selfadjoint operator \eqref{buslaev} has the form
\bead
\lefteqn{
\widetilde{L}(f \oplus g_l \oplus g_r)=}\\
& &
\begin{pmatrix}
-\frac{1}{2}\frac{d}{dx}\frac{1}{m(x)}\frac{d}{dx}f+ vf & 0 & 0 \\
0 & -\frac{1}{2}\frac{d}{dx}\frac{1}{m_l}\frac{d}{dx}g_l+v_lg_l & 0 \\
0 & 0 & -\frac{1}{2}\frac{d}{dx}\frac{1}{m_r}\frac{d}{dx}g_r+v_rg_r
\end{pmatrix}
\eead
and coincides with the usual Schr\"odinger operator 
\begin{equation*}
 -\frac{1}{2}\frac{d}{dx}\frac{1}{\widetilde m}\frac{d}{dx}+ 
\widetilde v\upharpoonright\left\{f \in L^2(\dR): f,\, 
\frac{1}{\widetilde m}f' \in W^{1,2}(\bR)\right\},
\end{equation*}
where
\bed
\widetilde m(x) := \left\{
\ba{ll}
m(x), & x \in (x_l,x_r)\\
m_l(x), & x \in (-\infty,x_l)\\
m_r(x), & x \in (x_r,\infty)
\ea
\right.
\eed
and
\bed
\widetilde v(x) := \left\{
\ba{ll}
v(x), & x \in (x_l,x_r)\\
v_l(x), & x \in (-\infty,x_l)\\
v_r(x), & x \in (x_r,\infty).
\ea
\right.
\eed

The selfadjoint operator $L_0=A_0\oplus T_0$, where $A_0=A^*\upharpoonright\ker(\Gamma_0)$ and 
$T_0=T^*\upharpoonright\ker(\gY_0)$, 
is defined on
\bead
\lefteqn{
\dom(L_0) = }\\
& & \hspace{-2mm}
\left\{f \oplus g_l \oplus g_r \in \dom(A^*) \oplus
\dom(T^*_l) \oplus \dom(T^*_r): 
\ba{l}f(x_l) = f(x_r) = 0\\
      g_l(x_l) = g_r(x_r) = 0
\ea
\right\}
\nonumber
\eead
and can be identified with the
selfadjoint Schr\"odinger operator 
\bed
-\frac{1}{2}\frac{d}{dx}\frac{1}{\widetilde m}\frac{d}{dx}+ \widetilde v\upharpoonright
\left\{
f \in L^2(\dR) :  
f,\frac{1}{\widetilde m}f' \in W^{1,2}(\bR \setminus \{x_l,x_r\})
\right\}.
\eed

\subsection{$S$ and $R$-matrix representation}

It is well known that all selfadjoint extensions of the differential 
operator $A$ in $L^2((x_l,x_r))$ have discrete
spectrum. Hence
according to Theorem~\ref{srzthm} and Corollary~\ref{srzthmcor} the selfadjoint Schr\"{o}dinger operators
$\widetilde L$ and $L_0$ form a complete scattering system $\{\widetilde L,L_0\}$ in $L^2(\dR)$ and
the scattering matrix $\{S(\gl)\}$ is given by
\begin{equation}\la{4.15}
S (\gl) = I_{\kH_{\tau(\lambda)}} -
2i\sqrt{\imag(\tau(\gl))}\bigl(M(\gl)+\tau(\gl)\bigr)^{-1}
\sqrt{\imag(\tau(\gl))}
\end{equation}
for $\gl \in \gS^M\cap \gS^\gt \cap \gS^{(M+\tau)^{-1}}$. Here $M(\cdot)$ is the Weyl function corresponding
to the boundary triplet $\Pi_A=\{\dC^2,\Gamma_0,\Gamma_1\}$ and 
\begin{equation*}
\lambda\mapsto\tau(\lambda)=\begin{pmatrix}\mathfrak m_l(\lambda) & 0\\ 0 & \mathfrak m_r(\lambda)
\end{pmatrix},\qquad\lambda\in\rho(T_0),
\end{equation*}
is the Weyl function of $\Pi_T=\{\dC^2,\gY_0,\gY_1\}$, cf. Section~\ref{2.3.2}.
It follows from \cite{KNR1} that for $\lambda\in\gS^\gt$ with  $\imag(\gt(\lambda)) \not= 0$
the maximal dissipative differential operator 
\begin{equation*}
A_{-\tau(\lambda)}=A^*\upharpoonright\ker\bigl(\Gamma_1+\tau(\lambda)\Gamma_0\bigr),
\end{equation*} 
that is,
\begin{equation*}
\begin{split}
\bigl(A_{-\tau(\lambda)}f\bigr)(x) &=
-\frac{1}{2}\frac{d}{dx}\frac{1}{m(x)}\frac{d}{dx}f(x) + v(x)f(x),\\
\dom(A_{-\tau(\lambda)}) &= \left\{f \in L^2 ((x_l,x_r)): 
\ba{c}
f, \frac{1}{m}f' \in W^{1,2}((x_l,x_r))\\
\left(\frac{1}{2m}f'\right)(x_l) = - \mathfrak m_l(\gl)f(x_l)\\
\left(\frac{1}{2m}f'\right)(x_r) =  \mathfrak m_r(\gl) f(x_r)
\ea
\right\},
\end{split}
\end{equation*}
has no real eigenvalues, i.e. $\dR\subset\rho(A_{-\tau(\lambda)})$, so that 
each $\lambda\in\Sigma^M=\rho(A_0)\cap\dR$
necessarily belongs to the set $\gS^{(M+\tau)^{-1}}$ by Theorem~\ref{resthm}. Therefore the representation 
\eqref{4.15} is valid for all 
$\lambda\in\{t \in \gS^\gt: \imag(\gt(t)) \not= 0\}\cap\rho(A_0)$. Moreover, for $\lambda\in\Sigma^\tau$ with
$\imag(\gt(\lambda))=0$ we have $S(\lambda)=\{0\}$.

It is well known that the symmetric operator $A$ given by \eqref{inner1} is semi-bounded
from below and that the extension $A_0 =
A^*\upharpoonright\ker(\gG_0)$, cf. \eqref{2.14}, is the 
Friedrichs extension of $A$. In particular, this yields $A_\gT \le A_0$ for any
other selfadjoint extension $A_\gT$ of $A$.

The selfadjoint operator $A_{-\real(\tau(\lambda))}$, $\gl \in \gS^\gt = 
\gS^{\mathfrak m_l} \cap \gS^{\mathfrak m_r}$, is given by
\begin{equation*}
\begin{split}
\bigl(A_{-\real(\tau(\lambda))}f\bigr)(x) &=
-\frac{1}{2}\frac{d}{dx}\frac{1}{m(x)}\frac{d}{dx}f(x) + v(x)f(x),\\
\dom(A_{-\real(\tau(\lambda))}) &= \left\{f \in L^2 ((x_l,x_r)): 
\ba{c}
f, \frac{1}{m}f' \in W^{1,2}((x_l,x_r))\\
\left(\frac{1}{2m}f'\right)(x_l) = - \real(\mathfrak m_l(\gl))f(x_l)\\
\left(\frac{1}{2m}f'\right)(x_r) = \real(\mathfrak m_r(\gl)) f(x_r)
\ea
\right\}
\end{split}
\end{equation*}
and clearly $\sigma(A_{-\real(\tau(\lambda))})$ is discrete and semi-bounded from
below for all $\gl \in \gS^\gt$. 

Taking into account Theorem~\ref{proppo} it follows that the $R$-matrix
of $\{\widetilde L,L_0\}$ has the form
\begin{equation*}
\begin{split}
R(\lambda)&=\sum_{k=1}^\infty(\lambda_k[\lambda]-\lambda)^{-1}
\left(\begin{pmatrix}\sqrt{\imag(\mathfrak m_l(\lambda))}\,\cdot\\ \sqrt{\imag(\mathfrak m_r(\lambda))}\,\cdot
 \end{pmatrix},\begin{pmatrix}\psi_k[\lambda](x_l)\\\psi_k[\lambda](x_r) \end{pmatrix}\right)\\
 &\qquad\qquad\qquad\qquad\qquad\qquad\qquad\qquad\qquad\cdot
\begin{pmatrix}\sqrt{\imag(\mathfrak m_l(\lambda))}\psi_k[\lambda](x_l) \\ 
\sqrt{\imag(\mathfrak m_r(\lambda))} \psi_k[\lambda](x_r)
 \end{pmatrix}
 \end{split}
\end{equation*}
for all $\lambda\in\Sigma^\tau\cap\Sigma^M$ with the property $\ker(M(\lambda)+\real(\tau(\lambda)))=\{0\}$ 
and $\imag(\tau(\lambda))\not=0$. Here $\{\lambda_k[\lambda]\}$, $k=1,2,\dots$, 
denote the eigenvalues of the selfadjoint
operator $A_{-\real(\tau(\lambda))}$ in increasing order and $\psi_k[\lambda]$ are the corresponding
eigenfunctions. Furthermore
we have again used $\dR\subset\rho(A_{-\tau(\lambda)})$ if $\imag(\tau(\lambda))\not=0$, and moreover, 
$R(\lambda)=\{0\}$ if $\imag(\tau(\lambda))=0$.

The scattering matrix $\{S(\lambda)\}$ of $\{\widetilde L,L_0\}$ can be represented in the form
\begin{equation*}
\begin{split}
S(\lambda)=&\Biggl\{iI_{\cH_{\tau(\lambda)}}-
\sum_{k=1}^\infty(\lambda_k[\lambda]-\lambda)^{-1}
\left(\begin{pmatrix}\sqrt{\imag(\mathfrak m_l(\lambda))}\,\cdot\\ \sqrt{\imag(\mathfrak m_r(\lambda))}\,\cdot
 \end{pmatrix},\begin{pmatrix}\psi_k[\lambda](x_l)\\\psi_k[\lambda](x_r) \end{pmatrix}\right)\\
&\qquad\qquad\qquad\qquad\qquad\qquad\qquad\qquad\qquad\cdot 
\begin{pmatrix}\sqrt{\imag(\mathfrak m_l(\lambda))}\psi_k[\lambda](x_l) \\ 
\sqrt{\imag(\mathfrak m_r(\lambda))} \psi_k[\lambda](x_r)
 \end{pmatrix}\Biggr\}\\
& \times\, \Biggl\{iI_{\cH_{\tau(\lambda)}}+
\sum_{k=1}^\infty(\lambda_k[\lambda]-\lambda)^{-1}
\left(\begin{pmatrix}\sqrt{\imag(\mathfrak m_l(\lambda))}\,\cdot\\ \sqrt{\imag(\mathfrak m_r(\lambda))}\,\cdot
 \end{pmatrix},\begin{pmatrix}\psi_k[\lambda](x_l)\\\psi_k[\lambda](x_r) \end{pmatrix}\right)\\
&\qquad\quad\qquad\qquad\qquad\qquad\qquad\qquad\qquad \cdot
\begin{pmatrix}\sqrt{\imag(\mathfrak m_l(\lambda))}\psi_k[\lambda](x_l) \\ 
\sqrt{\imag(\mathfrak m_r(\lambda))} \psi_k[\lambda](x_r)
 \end{pmatrix}\Biggr\}^{-1}
\end{split}
\end{equation*}
for all $\lambda\in\Sigma^\tau\cap\Sigma^M$ with $\ker(M(\lambda)+\real(\tau(\lambda)))=\{0\}$ 
and $\imag(\tau(\lambda))\not=0$.

\subsubsection{Constant potentials $v_l$ and $v_r$}\label{vlvr}

Let us assume that the potentials  $v_l(\cdot)$ and $v_r(\cdot)$ as well as
the mass functions $m_l(\cdot)$ and $m_r(\cdot)$ are constant,
that is, $v_l(x) = v_l\in\dR$, $m_l(x) = m_l>0$ for $x \in (-\infty,x_l)$ and
$v_r(x) = v_r\in\dR$, $m_r(x) = m_r>0$ for $x \in (v_r,\infty)$.
The Titchmarsh-Weyl functions $\mathfrak m_l(\cdot)$ and  $\mathfrak m_r(\cdot)$ can be
calculated explicitly in this simple case, see \cite{BMN2}. One gets
\bed
\mathfrak m_l(\gl) = i\sqrt{\frac{\gl - v_l}{2m_l}} 
\qquad \mbox{and} \qquad
\mathfrak m_r(\gl) = i\sqrt{\frac{\gl - v_r}{2m_r}} 
\eed
for $\gl \in \bC_+$,
where the square root is defined on $\bC$ with a cut along $[0,\infty)$ and fixed by
$\imag(\sqrt{\gl})>0$ for $\lambda\not\in [0,\infty)$ and by $\sqrt{\lambda}\geq 0$ for
$\lambda\in[0,\infty)$. It is clear that
\bed
\gS^{\gt} = \gS^{\mathfrak m_l}\cap\gS^{\mathfrak m_r}  = \bR
\eed
and it is not difficult to check
\bed
\bigl\{\gl \in \gS^\gt: \imag(\gt(\gl)) \not= 0\bigr\} = \bigl(\min\{v_l,v_r\},\infty\bigr).
\eed
Furthermore
\begin{equation*}
\real(\mathfrak m_l(\gl)) = 
\begin{cases}
-\sqrt{\frac{v_l - \gl}{2m_l}}, & \gl \le v_l,\\
0, & \gl > v_l,
\end{cases} 
\end{equation*}
and
\bed
\real(\mathfrak m_r(\gl)) = 
\begin{cases}
-\sqrt{\frac{v_r - \gl}{2m_r}}, & \gl \le v_r,\\
0, & \gl > v_r.
\end{cases} 
\eed
If $\gl \in (\max\{v_l,v_r\},\infty)$, then $\real(\gt(\gl)) =
0$ and it follows from Corollary~\ref{0cor} and the above considerations 
that the $R$-matrix of $\{\widetilde L,L_0\}$ has the form
\begin{equation}\label{exdiv}
\begin{split}
R(\lambda)&=\sum_{k=1}^\infty(\lambda_k-\lambda)^{-1}
\left(\begin{pmatrix}\sqrt{\imag(\mathfrak m_l(\lambda))}\,\cdot\\ \sqrt{\imag(\mathfrak m_r(\lambda))}\,\cdot
 \end{pmatrix},\begin{pmatrix}\psi_k(x_l)\\\psi_k(x_r) \end{pmatrix}\right)\\
 &\qquad\qquad\qquad\qquad\qquad\qquad\qquad\qquad\qquad\cdot
\begin{pmatrix}\sqrt{\imag(\mathfrak m_l(\lambda))}\psi_k(x_l) \\ 
\sqrt{\imag(\mathfrak m_r(\lambda))} \psi_k(x_r)
 \end{pmatrix}
 \end{split}
\end{equation}
for all $\lambda\in\Sigma^M$ with the property $\ker(M(\lambda))=\{0\}$. 
Here $\{\lambda_k\}$, $k=1,2,\dots$, 
denote the eigenvalues of the selfadjoint
operator $A_1$ in increasing order and $\psi_k$ are the corresponding
eigenfunctions. Note that $A_1$ is the usual Schr\"{o}dinger operator in $L^2((x_l,x_r))$ 
which corresponds to Neumann boundary
conditions, cf. \eqref{2.33b}, and that $\lambda\in\Sigma^M$ has the property $\ker(M(\lambda))=\{0\}$
if and only if $\lambda\in\rho(A_0)\cap\rho(A_1)$, cf. Theorem~\ref{resthm}. 

Analogously the scattering matrix $\{S(\lambda)\}$ of $\{\widetilde L,L_0\}$ has the form
\begin{equation*}
\begin{split}
S(\lambda)=&\Biggl\{iI_{\cH_{\tau(\lambda)}}-
\sum_{k=1}^\infty(\lambda_k-\lambda)^{-1}
\left(\begin{pmatrix}\sqrt{\imag(\mathfrak m_l(\lambda))}\,\cdot\\ \sqrt{\imag(\mathfrak m_r(\lambda))}\,\cdot
 \end{pmatrix},\begin{pmatrix}\psi_k(x_l)\\\psi_k(x_r) \end{pmatrix}\right)\\
&\qquad\qquad\qquad\qquad\qquad\qquad\qquad\qquad\qquad\cdot 
\begin{pmatrix}\sqrt{\imag(\mathfrak m_l(\lambda))}
\psi_k(x_l) \\ 
\sqrt{\imag(\mathfrak m_r(\lambda))} \psi_k(x_r)
 \end{pmatrix}\Biggr\}\\
& \times\, \Biggl\{iI_{\cH_{\tau(\lambda)}}+
\sum_{k=1}^\infty(\lambda_k-\lambda)^{-1}
\left(\begin{pmatrix}\sqrt{\imag(\mathfrak m_l(\lambda))}\,\cdot\\ \sqrt{\imag(\mathfrak m_r(\lambda))}\,\cdot
 \end{pmatrix},\begin{pmatrix}\psi_k(x_l)\\\psi_k(x_r) \end{pmatrix}\right)\\
&\qquad\quad\qquad\qquad\qquad\qquad\qquad\qquad\qquad \cdot
\begin{pmatrix}\sqrt{\imag(\mathfrak m_l(\lambda))}
\psi_k(x_l) \\ 
\sqrt{\imag(\mathfrak m_r(\lambda))} \psi_k(x_r)
 \end{pmatrix}\Biggr\}^{-1}
\end{split}
\end{equation*}
for all $\gl \in (\max\{v_l,v_r\},\infty)\cap\rho(A_0)\cap\rho(A_1)$.

The situation is slightly more complicated if $\gl \in (\min\{v_l,v_r\},\max\{v_l,v_r\})$.
Assume e.g. $v_l > v_r$ and let $\lambda\in(v_r,v_l)$. In this case $\imag(\tau(\lambda))\not=0$, 
but the condition $\real(\tau(\gl))
= 0$ is not satisfied since 
\bed
\real(\mathfrak m_l(\gl)) = -\sqrt{\frac{v_l - \gl}{2m_l}} 
\quad \mbox{and} \quad
\real(\mathfrak m_r(\gl)) = 0.
\eed
The operator
$A_{-\real(\tau(\gl))}$ is given by
\begin{equation*}
\begin{split}
\bigl(A_{-\real(\tau(\lambda))}f\bigr)(x) &=
-\frac{1}{2}\frac{d}{dx}\frac{1}{m(x)}\frac{d}{dx}f(x) + v(x)f(x),\\
\dom(A_{-\real(\tau(\lambda))}) &= \left\{f \in L^2 ((x_l,x_r)): 
\ba{c}
f, \frac{1}{m}f' \in W^{1,2}((x_l,x_r))\\
\left(\frac{1}{2m}f'\right)(x_l) =  \sqrt{\frac{v_l - \gl}{2m_l}} f(x_l)\\
\left(\frac{1}{2m}f'\right)(x_r) = 0
\ea
\right\}.
\end{split}
\end{equation*}

Since
\bed
\sqrt{\imag(\gt(\gl))} =
\begin{pmatrix}
0 & 0\\
0 & \left(\frac{\gl - v_r}{2m_r}\right)^{1/4}
\end{pmatrix},\qquad \gl  \in (v_r,v_l),
\eed
the representations of the $R$ and $S$-matrix of $\{\widetilde L,L_0\}$ from the previous subsections
become  
\begin{equation*}
\begin{split}
R(\lambda)&=\sum_{k=1}^\infty(\lambda_k[\lambda]-\lambda)^{-1}
\bigl(\sqrt{\imag(\mathfrak m_r(\lambda))}\,\cdot,\psi_k[\lambda](x_r) \bigr) 
\sqrt{\imag(\mathfrak m_r(\lambda))} \psi_k[\lambda](x_r)
 \end{split}
\end{equation*}
and
{\small
\begin{equation*}
S(\lambda)=\frac{i-\sum_{k=1}^\infty(\lambda_k[\lambda]-\lambda)^{-1}
\bigl(\sqrt{\imag(\mathfrak m_r(\lambda))}\,\cdot,\psi_k[\lambda](x_r) \bigr) 
\sqrt{\imag(\mathfrak m_r(\lambda))} \psi_k[\lambda](x_r)}
{i+\sum_{k=1}^\infty(\lambda_k[\lambda]-\lambda)^{-1}
\bigl(\sqrt{\imag(\mathfrak m_r(\lambda))}\,\cdot,\psi_k[\lambda](x_r) \bigr) 
\sqrt{\imag(\mathfrak m_r(\lambda))} \psi_k[\lambda](x_r)},
\end{equation*}
}
respectively, 
for $\gl \in (v_r,v_l) \cap \rho(A_0) \cap
\rho(A_{-\real(\tau(\gl))})$, see Theorem~\ref{proppo}. 
Here $\{\lambda_k[\lambda]\}$, $k=1,2,\dots$, are the eigenvalues of the selfadjoint extension 
$A_{-\real(\tau(\lambda))}$ in increasing
order and $\psi_k[\lambda]$ are the corresponding eigenfunctions.

\begin{rem}
{\rm One might guess that the sum 
\bed
\sum^\infty_{k=1}(\gl_k - \gl)^{-1}\left(\cdot,
\begin{pmatrix}
\psi_k(x_l)\\
\psi_k(x_r)
\end{pmatrix}
\right)
\begin{pmatrix}
\psi_k(x_l)\\
\psi_k(x_r)
\end{pmatrix}
\eed
in the representation of the scattering matrix in \eqref{exdiv}, 
where $\{\gl_k\}$ and $\{\psi_k\}$ are the eigenvalues and eigenfunctions
of the Schr\"odinger operator with Neumann boundary conditions, can be replaced
by the sum
\bed
\sum^\infty_{k=1}(\mu_k - \gl)^{-1}\left(\cdot,
\begin{pmatrix}
(\frac{1}{2m}\phi'_k)(x_l)\\
-(\frac{1}{2m}\phi'_k)(x_r)
\end{pmatrix}
\right)
\begin{pmatrix}
(\frac{1}{2m}\phi'_k)(x_l)\\
-(\frac{1}{2m}\phi'_k)(x_r)
\end{pmatrix},
\eed
where $\{\mu_k\}$ and $\{\phi_k\}$ are the eigenvalues and
eigenfunctions of the Schr\"odinger operator with Dirichlet boundary
conditions. However, this is not possible since by Proposition
\ref{negativ} the last sum does not converge. We note that this can easily be
verified by hand for the case $v(x) = 0$ and $m(x) = \mbox{constant}$.
}
\end{rem}

\end{document}